\documentclass[graphicx,11pt]{aastex}

\lefthead{}
\righthead{}

\begin{document}

\title{The great dichotomy of the Solar System: small terrestrial embryos and massive giant planet cores} 

\author{\textbf{A. Morbidelli$^{(1)}$,M. Lambrechts$^{(2)}$, S. Jacobson$^{(1,3)}$, B. Bitsch$^{(2)}$}\\  
(1) Laboratoire Lagrange, UMR7293, Universit\'e de Nice Sophia-Antipolis,
  CNRS, Observatoire de la C\^ote d'Azur. Boulevard de l'Observatoire,
  06304 Nice Cedex 4, France. (Email: morby@oca.eu / Fax:
  +33-4-92003118) \\
(2) Lund Observatory, Department of Astronomy and Theoretical Physics, Lund University, Box 43, 22100, Lund, Sweden  \\ 
(3) Bayerisches Geoinstitut, University of Bayreuth, D-95440 Bayreuth, Germany\\
} 

\begin{abstract}

The basic structure of the solar system is set by the presence of low-mass terrestrial planets in its inner part and giant planets in its outer part. This is the result of the formation of a system of multiple embryos with approximately the mass of Mars in the inner disk and of a few multi-Earth-mass cores in the outer disk, within the lifetime of the gaseous component of the protoplanetary disk. What was the origin of this dichotomy in the mass distribution of embryos/cores? We show in this paper that the classic processes of runaway and oligarchic growth from a disk of planetesimals cannot explain this dichotomy, even if the original surface density of solids increased at the snowline. Instead, the accretion of drifting pebbles by embryos and cores can explain the dichotomy, provided that some assumptions hold true. We propose that the mass-flow of pebbles is two-times lower and the characteristic size of the pebbles is approximately ten times smaller within the snowline than beyond the snowline (respectively at heliocentric distance $r<r_{ice}$ and $r>r_{ice}$, where $r_{ice}$ is the snowline heliocentric distance), due to ice sublimation and the splitting of icy pebbles into a collection of chondrule-size silicate grains. In this case,  objects of original sub-lunar mass would grow at drastically different rates in the two regions of the disk. Within the snowline these bodies would reach approximately the mass of Mars while beyond the snowline they would grow to $\sim 20$ Earth masses. The results may change quantitatively with changes to the assumed parameters, but the establishment of a clear dichotomy in the mass distribution of protoplanets appears robust {provided that there is enough turbulence in the disk to prevent the sedimentation of the silicate grains into a very thin layer.} 

\end{abstract}

\section{Introduction}

The Solar System has a characteristic structure, with low-mass rocky planets in its inner part, often called terrestrial planets, and giant planets (gas-dominated or ice-dominated) in the outer part. 

A census of protoplanetary disks in clusters with known ages shows that the dust emission (usually assumed to trace the abundance of gas) disappears in a few My (Haisch et al., 2001); this is also the timescale on which the emission lines diagnostic of gas accretion onto the central star fade away (Hartmann et al., 1998). The fact that no primitive chondrite parent bodies seem to have accreted beyond 3--4 My (Kleine et al., 2005) suggests that the proto-Solar-System disk was not of exceptional longevity.

Clearly, the giant planets had to form within the lifetime of the gas-disk because they accreted  substantial amounts of hydrogen and helium (this is true also for Uranus and Neptune). Thus giant planets should have formed within a few My only. The commonly accepted scenario for giant-planet formation is the core-accretion model (Pollack et al., 1996). In short, a massive solid core accretes first and then it captures a massive atmosphere of H and He from the protoplanetary disk. The mass of all giant planet cores but Jupiter is around 10 Earth masses ($M_\oplus$) (Guillot, 2005). Jupiter might have no core today (Nettelmann et al., 2008). However, there are several tens of Earth masses of ``metals'' (molecules heavier than H and He) in Jupiter (Guillot, 2005) and it is possible that part or even most of its primordial core has been eroded and dissolved into the atmosphere (Guillot et al., 2004; Wilson and Militzer, 2012). 

An estimate of the mass of the core needed for the accretion of a massive atmosphere is also provided by models. It is generally considered, since the work of Pollack et al., that the core needs to exceed $\sim 10 M_\oplus$. More precisely, the critical mass for the runaway accretion of the atmosphere depends on the rate at which the core accretes solids, on the molecular weight of the atmosphere (Ikoma et al., 2000; Hori and Ikoma, 2011) and the dust opacity in the envelope (Mizuno, 1980, Stevenson, 1982), {which remains poorly known despite modern attempts to estimate the dust opacity self-consistently by modeling the aggregation of infalling grains (Movshovitz and Podolak, 2008; Ormel, 2014)}. The solid accretion rate could not be arbitrarily small, otherwise the core would not have formed in first place within the lifetime of the disk. This gives a constraint on the minimal mass of the core. Lambrechts et al. (2014; see their Fig. 7) showed that the core should have had a mass of at least 10 $M_\oplus$, if the ratio H$_2$O/H$_2$ in the atmosphere was less than 0.6.  Uranus and Neptune provide an indirect confirmation of this estimate as a lower-bound for the core mass. In fact, they have a core of about 10-15 $M_\oplus$ (Guillot, 2005) and only a few Earth masses of H and He, which means that they either did not start runaway accretion of gas, or did so only at the very end of the lifetime of the disk. 

The situation for the terrestrial planets is completely different. There is a general consensus that the terrestrial planets formed from a system of planetary embryos and planetesimals (see Morbidelli et al., 2012 for a review), although the details of how this happened can differ from one model to the other (Chambers and Wetherill, 1998; Chambers, 2001; Agnor et al., 1999; Raymond et al., 2004, 2006; O'brien et al., 2006; Hansen, 2009; Walsh et al., 2011; Jacobson and Morbidelli, 2014). According to these models and to the interpretation of isotopic chronometers for terrestrial and lunar samples (Yin et al., 2002; Jacobsen, 2005; Touboul et al., 2007; Allegre et al., 2008; Halliday, 2008; Taylor et al., 2009) the Earth took several tens of My to complete its formation,  with a preferred timing for the Moon forming event around 100~My (Jacobson et al., 2014). The minimum time in which the Earth acquired 63\% of its mass is 11~My (Yin et al., 2002; Jacobsen, 2005). Thus, most of the assemblage of the Earth clearly took place after the removal of the gas from the protoplanetary disk. 

Mars, instead, formed very quickly, i.e. in a few My (Halliday and Kleine, 2006; Dauphas and Pourmand, 2011), basically on the same timescale of chondritic parent bodies. This suggests that Mars is a stranded embryo (Jacobson and Morbidelli, 2014). The fact that the Moon-forming projectile also had a mass of the order of a Mars-mass (Canup and Asphaug, 2001; Cuk and Stewart, 2012) {and that Mercury, if it had originally the same iron content as the other terrestrial planets, was also approximately Mars-mass (Benz et al., 1988)} suggests that the mass of Mars was the typical mass of planetary embryos in the inner Solar System at the time the gas was removed from the protoplanetary disk.

In summary, it appears compelling that, by the time gas was removed from the system, the process of formation of the solid component of planets had produced a great dichotomy in the mass distribution of protoplanets: in the inner system, the largest objects were approximately Mars-mass; in the outer solar system they were $\sim 10 M_\oplus$. Thus, there was a contrast of two orders of magnitude between the masses of the solid planets formed in the inner and outer systems respectively. This happened despite the accretion timescale, which can be reasonably approximated by the orbital timescale, is 10 times faster at 1 AU than 5 AU!

How could this be possible? The generic (and hand--waving) explanation is that the cores of the giant planets formed beyond the ice line, so that the density of solids was comparatively larger than in the inner solar system, where only refractory material could be in solid form. However, this can not be the explanation. According to Lodders (2003), for the solar abundance the H$_2$O-ice/rock ratio is approximately one-to-one. That means that the amount of solid mass available for planet formation beyond the snowline increases by just a factor of 2. This is confirmed by the ice/rock ratio inferred for comets, trans-Neptunian objects, and irregular satellites of giant planets (McDonnell et al. 1987; Stern, 1997; Johnson and Lunine, 2005). An enhancement of the solid mass by more than a factor of 2 might have been produced by the so-called ``cold-finger effect'' (Morfill and Volk, 1984; Ros and Johansen, 2013), but this would have happened only locally at the snowline and therefore could explain at most the formation of one giant-planet core, not several. 

The purpose of this paper is to investigate which process of planet formation is more likely to have led to the dichotomy discussed above. In section~2 we consider the classical process of formation of embryos/cores by runaway/oligarchic accretion of planetesimals (Greenberg et al., 1978; Kokubo and Ida, 1998; Wetherill and Stewart, 1993; Weidenschilling et al., 1997). We show that this process clearly cannot explain the dichotomy. Next, in section~3, we consider the process of pebble accretion. This is a new process for planet growth, introduced in Lambrechts and Johansen (2012; see also the precursor work by Ormel and Klahr, 2010; Johansen and Lacerda, 2010; Murray-Clay et al., 2011; Bromley and Kenyon, 2011), which is rapidly gaining attention (Morbidelli and Nesvorny, 2012; Chambers, 2014; Lambrechts and Johansen, 2014; Lambrechts et al., 2014; Guillot et al., 2014; Kretke and Levison, 2014a,b). We will show that, provided some assumptions hold true, the pebble accretion process can explain the two orders of magnitude mass-contrast between inner solar system objects and outer solar system objects. The conclusions and perspectives are discussed in section~4. 

\section{Growth of embryos and cores by planetesimal accretion}

The growth of embryos and cores from a disk of planetesimals proceeds in two phases. 

The first phase is that of runaway growth (Greenberg et al., 1978). Here most of the mass of the disk is in ``small'' planetesimals. The velocity dispersion of the planetesimals is set by the equilibrium between the self-excitation of their orbits, also called self-stirring, and gas drag. We neglect here collisional damping because it is important only for very small objects, which we will call pebbles in the next section, and in the absence of gas drag (Goldreich et al., 2004; Levison and Morbidelli, 2007). The velocity dispersion of the planetesimals is therefore comparable or smaller (because of the drag) to the escape velocity from the surface of the planetesimals carrying the bulk of the population mass (Greenberg et al., 1978). In this situation, the accretion cross section $\sigma$ of an individual planetesimal is: 
\begin{equation}
\sigma=\pi R^2 \left(1+{{V_{esc}}^2\over{V_{rel}}^2}\right)\ ,
\label{focussing}
\end{equation}
where $R$ is the planetesimal radius, $V_{esc}$ is the escape velocity from the planetesimal surface, $V_{rel}$ is the dispersion velocity in the planetesimal disk and the term in parenthesis is called the  ``gravitational focusing factor'' (Greenberg et al., 1978; Greenzweig and Lissauer, 1990, 1992). Thus, the most massive planetesimals have a comparative advantage. If their $V_{esc}$ is significantly larger than $V_{rel}$, their gravitational focusing factor can be approximated by $V_{esc}^2/V_{rel}^2\propto M^{2/3}/V_{rel}^2$, where $M$ is their mass. Because the accretion rate is proportional to $\sigma$ and $R\propto M^{1/3}$, from (\ref{focussing}) the relative mass accretion rate of the massive bodies is: 
\begin{equation}
{{1}\over{M}}{{{\rm d}M}\over{{\rm d}t}}\propto M^{1/3}\ .
\label{runaway}
\end{equation}
Eq. (\ref{runaway}) means that the most massive bodies grow the fastest and their mass ratio with the rest of the planetesimal population increases exponentially with time. Hence the name ``runaway growth''.

The second phase is that of oligarchic growth (Kokubo and Ida, 1998). It occurs when the runaway bodies become massive enough to stir-up the velocity dispersion of the planetesimals to a value comparable to the escape speed of the runaway bodies. In this case the gravitational focusing factor becomes of order unity and 
therefore eq.~(\ref{focussing}) becomes:
\begin{equation}
{{1}\over{M}}{{{\rm d}M}\over{{\rm d}t}}\propto M^{-1/3}\ .
\label{oligarchic}
\end{equation}
This equation means that the more massive a body is, the slower (in relative terms) it grows. The difference in mass between bodies becomes smaller and smaller as time passes. In principle, the planetesimals could reduce the mass difference with the former runaway bodies. But in general, when (\ref{oligarchic}) holds, the relative velocities are so large that collisions among the planetesimals become disruptive. Thus, only the massive bodies keep growing, forming a population of oligarchs that dominate the dynamical evolution in the disk. Hence the name ``oligarchic growth''. Numerical simulations (Kokubo and Ida, 2000) show that the oligarchs accommodate their orbits (possibly by merging with each other) so that their separation in semi major axis is 5-10 mutual Hill radii. Oligarchic growth stops when an oligarch has accreted most of the mass available in the annulus around its orbit that extends up to half-way to the neighboring oligarch.  

In order to compare the masses of protoplanets growing in different parts of the protoplanetary disk, below we first compare the initial runaway growth rates, then the final masses at the end of the oligarchic growth phase.

\subsection{Radial dependence of the runaway growth rate}
\label{disk}

Here we assume that, at the beginning of the growth phase, both the planetesimals and the runaway bodies have the same mass in all parts of the disk. This is in line with the classic assumptions that planetesimals are km-size everywhere in the disk and the runaway bodies are initially barely bigger than the characteristic planetesimal size (e.g. Kokubo and Ida, 1996; Greenberg et al., 1984). Under this assumption, on which we will return at the end of this section, the radial dependence of the growth rate of the runaway bodies is set by the radial dependencies of the spatial density of the planetesimal population and of the orbital properties (i.e. velocity dispersion). 

First, let's analyze how the spatial density of the planetesimal population $\rho_p$ depends on $r$. For this purpose we start from the radial profile of the gas density.

Whatever its initial distribution, the inner part of a protoplanetary disk of gas should rapidly evolve, under the action of viscosity, towards a distribution such that the mass-flow towards the star is independent of $r$ (Lynden-Bell and Pringle, 1974). This is called an ``accretion disk''. In this situation, and assuming the usual $\alpha$-prescription for the viscosity (Shakura and Sunyaev,  1973), the exponent $a$ of the power-law describing the density profile of the disk $\Sigma_g\propto 1/r^a$ is: $a=2{{\rm d}\log H_g\over{{\rm d}\log r}} -3/2$ (Bitsch et al., 2014a). Here $H_g$ is the pressure scale height of the gas. Bitsch et al. (2014a, 2015) showed that the outer part of the disk is usually flared under the effect of stellar irradiation, so that ${{\rm d}\log H_g\over{{\rm d}\log r}} = 9/7$ and $a=15/14$. However, the inner part of the disk, dominated by viscous heating, is not flared. Thus  ${{\rm d}\log H_g\over{{\rm d}\log r}} \sim 1$ (apart from some oscillations due to opacity transitions, which we neglect here). For an early disk, with a mass flow to the star of $10^{-7} M_\odot/y$ (where $M_\odot$ is the mass of the Sun), the transition between these two regimes occurs at $\sim 6$~AU. Thus, for our purposes we consider that the disk is not flared, so that $a\sim 1/2$. The volume density of the gas $\rho_g\propto \Sigma_g/H_g$ is therefore proportional to $1/r^{3/2}$. 

This disk's density profile is much shallower than that in the usual Minimum Mass Solar Nebula (MMSN) model (Weidenschilling, 1977; Hayashi, 1981), in which $\Sigma_g\propto 1/r^{3/2}$ and $\rho_g\propto 1/r^{5/2}$. 
The fundamental assumptions of the MMSN model (the planets formed in situ, exclusively from local material), however, are not considered valid any more (e.g. Crida, 2009) given what we now today know about planet migration; thus we think that the shallow disk profile with $a=1/2$ has a stronger theoretical motivation. It is also important to realize that if the disk's density profile is shallower, then the chances to form massive planets further out increases {(e.g. Lissauer, 1987)}. Thus, using a shallower disk is more favorable for explaining the solar system dichotomy. 

At the beginning of the accretion phase, it is commonly assumed that the surface density of solids $\Sigma_p$ is proportional to that of the gas: $\Sigma_p=m \Sigma_g$, where $m$ is the metallicity factor. As we said in the introduction $m$ should increase by a factor of $\sim 2$ at the snowline because of the presence of water ice (Lodders, 2003). The volume density of planetesimals on the disk's midplane, $\rho_p$, is proportional to $\Sigma_p/r\sin(i)$, where $i$ is the inclination dispersion of the planetesimals.  It is well-known that for a disk undergoing self-stirring, $i\propto e/2$ (Stewart and Ida, 2000). Thus we have:
\begin{equation}
\rho_p\propto {{m \rho_g}\over{e}} \propto {{m}\over{r^{3/2}e}}\ .
\label{rho}
\end{equation}

We now analyze the velocity dispersion of the planetesimal disk, namely its eccentricity excitation.

As we said above, the eccentricity excitation is set by the equilibrium between self-stirring of the planetesimal population and gas drag. 
For a disk of equal mass particles, the growth rate of the eccentricity dispersion is (Stewart and Ida, 2000; Morbidelli et al., 2009 - supplementary material):
\begin{equation}
{{{\rm d}e^2}\over{{\rm d}t}}\propto 
{{\rho_p M}\over{V_K^3 e}} \ ,  
\label{de2vs}
\end{equation}
where $M$ is the individual mass of the planetesimals and $V_K$ is the keplerian velocity. 

For the gas drag, the evolution of the eccentricity excitation is (Wetherill and Stewart, 1989; Morbidelli et al., 2009 - supplementary material):
\begin{equation}
{{{\rm d}e^2}\over{{\rm d}t}}\propto {{e\rho_g V_g^2 R^2}\over{V_K M}}
\label{drag}
\end{equation}
where $R$ is the radius of the planetesimals and $V_g$ is the orbital velocity of the gas which, for eccentric planetesimal orbits, can be safely approximated by $V_K$.

Thus, by equating (\ref{de2vs}) and (\ref{drag}) and using the relationship (\ref{rho}) and keeping only the terms which have a radial dependence, we find that the velocity dispersion is:
\begin{equation}
V_{rel}\equiv eV_K \propto {{1}\over{V_K^{1/3}}} \propto r^{1/6}\ ,
\label{dispersion}
\end{equation}
so that $e\propto r^{2/3}$.

We can now use these results in the accretion formula:
\begin{equation}
\dot{M}=\rho_p\sigma V_{rel}\ ,
\end{equation}
with $\sigma$ given in (\ref{focussing}). We find:
\begin{equation}
\dot{M}\propto {{m \rho_g}\over{e V_{rel}}}\propto m r^{-7/3} \ .
\label{accretion}
\end{equation}. 

{This accretion rate is the same as that found in eq.~7 of Thommes et al. (2003), if one remembers that $\Sigma_m/e_m$ in that paper is the same as $r\rho_p$ here and one plugs in the radial dependences of $\rho_p$ and $e$ given above.} 

The conclusion is that accretion is much faster in the inner disk than in the outer disk, even in a disk with the shallowest possible density radial profile like the one we considered. With the radial dependence given in (\ref{accretion}), even taking into account that $m$ is twice larger beyond the snowline, the accretion rate at 1 AU is 20 times faster than at 5 AU! Because in runaway growth the relative accretion rate $\dot{M}/M\propto M^{1/3}$, absorbing this factor of 20 between the accretion rates at the two locations would require that the runaway body at 5 AU is initially 8,000 times more massive than at 1 AU. We don't see any justification of why it should be so. 

\subsection{Radial dependence of the final mass of the oligarchs}

The completion of oligarchic growth is quite an idealized concept. Oligarchic growth is rather slow and therefore it takes time to bring the process to completion, easily exceeding the lifetime of the gas in the protoplanetary disk. The simulations of Levison et al. (2012), the most realistic to date as they combine embryos growth with planetesimal grinding and account for a full size-distribution of objects, show that oligarchic growth approaches completion only in the very inner part of the disk. This result is in agreement with the strong radial dependence of the runaway growth rate found in the previous section.  

Nevertheless, it is instructive to compute which kind of radial mass distribution of objects oligarchic growth would predict, if it reached completion. By definition of ``completion'', the mass of the body is of the order of that originally available in an annulus around the orbit with a width equal to $n$ times the Hill radius $R_H$ of the body itself, with $n\sim 10$, independent of heliocentric distance (Kokubo and Ida, 2000). Thus we can write
\begin{equation}
M_{oli}= 2\pi r n R_H \Sigma_p.
\label{oli}
\end{equation}
With the shallow $\Sigma_p\propto m/\sqrt{r}$ defined above and remembering that $R_H=r(M/3M_\odot)^{1/3}$, eq.~(\ref{oli}) gives $M_{oli}\propto m r^{9/4}$. 

\begin{figure}[t!]
\centerline{\includegraphics[height=8.cm]{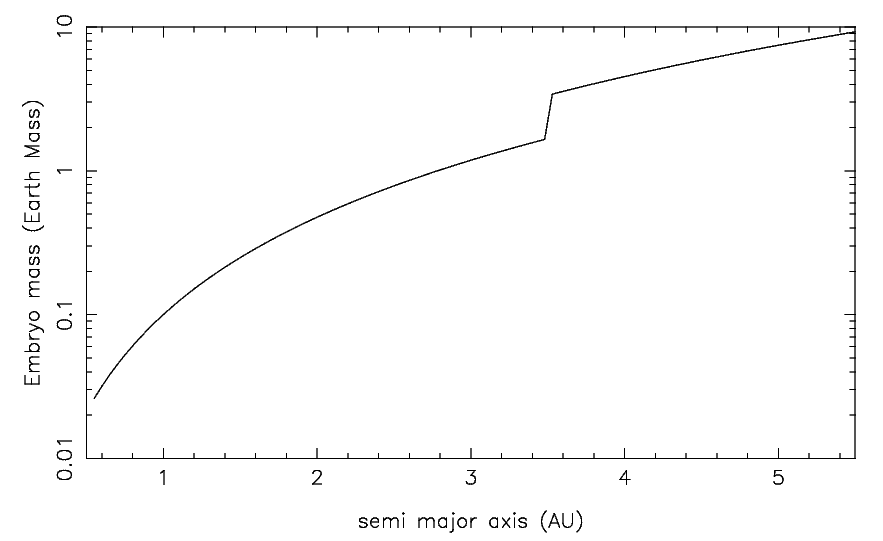}}
\caption{\small The mass distribution of embryos predicted by (\ref{oli}), for parameters such that the embryo at 1 AU has a 0.1~$M_\oplus$. The snowline is set here at 3.5 AU.}
\label{olifig}
\end{figure}

This is a strong growing function of $r$. If one accounts for an increase of $m$ by $\sim 2$ beyond the snowline, $M_{oli}$ predicts a ratio of almost 100 between the oligarchic masses at 5 AU and 1 AU! Is this the sign that oligarchic growth reached completion throughout the Solar System? 

We doubt that this is the answer for two reasons. First, look at Fig.~\ref{olifig}, which shows the mass distribution of the oligarchs predicted by the formula above. Here, the snowline effect has been assumed at 3.5~AU.  The formula does not predict Mars-mass embryos through the inner solar system and massive cores beyond the snowline. It predicts a mass distribution smoothly growing with distance up to the snowline, with Earth mass bodies in the asteroid belt. We don't see how this distribution of bodies could be reconciled with the current structure of the Solar System, given current models of terrestrial planet formation, planet migration etc. Second, Levison et al. (2010) showed that when bodies achieve a few Earth masses in the outer part of the disk their growth slows down enormously, because they start to scatter away the planetesimals from their feeding zone, rather than accreting them. Thus, all masses predicted in Fig.~\ref{olifig} beyond 3~AU are most likely overestimated and $\sim 10 M_\oplus$ cores are unable to form.

In summary, we conclude that the classical scheme of runaway/oligarchic growth of protoplanets from a disk of planetesimals can not explain the great dichotomy of the Solar System and hence its current structure. In the next section we turn to a new paradigm for planetary growth, that of pebble accretion (Lambrechts and Johansen, 2012).   

\section{Growth of embryos and cores by pebble accretion}
\label{formulae}

In a series of recent papers (Lambrechts \& Johansen, 2012, 2014; Ormel and Klahr, 2010; Johansen and Lacerda, 2010; Murray-Clay et al., 2011; Morbidelli and Nesvorny, 2012; Kretke and Levison, 2014b), a new
model has been proposed in which massive planetesimals accrete pebbles (particles cm to dm in size) very efficiently,
thanks to the combination of gravitational deflection and gas drag.
If most of the solid mass of the disk is in pebbles, the largest planetesimals can rapidly grow to several Earth masses.  

The process of pebble accretion can be illustrated and quantified as
follows.  First, it is convenient to define the {\it Bondi radius} $R_B$ as
the distance at which a planetesimal of mass $M$, assumed to be on a
circular orbit, exerts a deflection of one radian on a particle
approaching with a speed $\Delta v$ equal to the relative speed of the
gas\footnote{Notice that this definition of the Bondi radius, given in Lambrechts and Johansen (2012) is different from the classic definition of Bondi radius given in studies of bound planetary atmospheres, namely the distance from the planet at which the escape speed is equal to the sound speed.} . Because the gas is pressure supported, its orbital velocity is sub-Keplerian, and therefore it approaches the planetesimal with a speed $\Delta v=\eta V_K$ with:
\begin{equation}
\eta=-{{1}\over{2}}\left(H\over r\right)^2 {{{\rm d}\log P}\over{{\rm
      d}\log r}}\ ,
\label{eta}
\end{equation}
where $P$ is the gas pressure. Denoting by $G$ the gravitational constant, one has:
\begin{equation}
R_B={{GM}\over{(\Delta v)^2}}\ .
\label{Bondi}
\end{equation}
Due to three-body effects, however, the deflection cannot be effective beyond the Hill radius $R_H\equiv r[M/(3M_\odot)]^{1/3}$ of the planetesimal. Thus, we define :
$$
R_{GP}=\min(R_B,R_H)\ ,
$$
as the distance from the planetesimal within which its gravitational pull is effective. { Because the Bondi radius is proportional to $M$ while the Hill radius is proportional to $M^{1/3}$, for small bodies $R_{GP}=R_B$, while for more massive bodies $R_{GP}=R_H$. For a typical value of $\eta$ of 0.0027--0.0032, one has $R_B=R_H$ for $M\sim 2$--3$\times 10^{-3} M_\oplus$. Here we will consider for simplicity only bodies more massive than this threshold, so we detail the process of pebble accretion only in the case $R_{GP}=R_H$.}

Lambrechts and Johansen (2012) find {analytically and numerically} that for pebbles with Stokes number $\tau$ 
(the product between the friction time -the timescale on which a particle looses by friction its velocity relative to the gas- and the orbital frequency) the effective radius for accretion onto the planetesimal is:
\begin{equation}
r_{eff}=(\tau/0.1)^{1/3} R_{GP}\ ,
\label{reff}
\end{equation}
{which is valid for $\tau\le 0.1$. Notice also that the formula above has been derived for accreting bodies on circular orbits, for which the velocity difference with the pebbles is due to the Keplerian shear. If the body is eccentric, its velocity relative to the circular orbit has to be considered in the calculation of the relative velocity of the pebbles and, when the relative velocity increases, the value of $r_{eff}$ drops. In Sect.~\ref{self} we will confirm that the eccentricity excitation of accreting embryos can be neglected.}

To compute the growth rate on a body, one needs to distinguish between 2D and 3D accretion, depending on the pebbles scale height $H_{pb}$ relative to $r_{eff}$. Below we review the basic formul\ae\ for 2D and 3D accretion, then decide what is the threshold for $H_{pb}$ to pass from one regime to the other one.

\vskip 12pt
\noindent {\bf 2D accretion}

Let's consider that $H_{pb}$ is small relative to $r_{eff}$. 
The accretion rate of the body is then:
$$
\dot{M}_{2D}=2 r_{eff} v_{rel}\Sigma_{pb}
$$
where $\Sigma_{pb}$ is the surface density of the disk of pebbles and $v_{rel}$ is the relative velocity between the pebble and the body, namely 
\begin{equation}
v_{rel}=\max(\Delta v,v_{shear})
\label{vrel}
\end{equation}
and $v_{shear}=r_{eff}\Omega$, $\Omega=\sqrt{GM_\odot/r^3}$. 
{Because pebbles undergo inward radial drift due to gas drag, there is a continuous mass flux $\dot{M}_F$ of pebbles through any orbital radius. Thus, the steady-state} surface density of pebbles is related to this mass radial flux via the relationship:
$$
\Sigma_{pb}=\dot{M}_F/  (2 \pi r v_r)
$$
where $v_r$ is the radial drift velocity of pebbles, which can be approximated for small $\tau$ as $v_r=2\tau \Delta v$.

Thus we have:
\begin{equation}
\dot{M}_{2D}={{2 r_{eff} v_{rel} \dot{M}_F}\over{4\pi r \tau \Delta v}} \ .
\label{Mdot2D}
\end{equation}

\vskip12pt
\noindent{\bf 3D accretion}

In this case we have:
\begin{equation}
\dot{M}_{3D}=\pi r_{eff}^2 v_{rel}\rho_{pb}
\label{3Daccrete}
\end{equation}
where $\rho_{pb}$ is the volume density of pebbles on the midplane, which is related to the surface density $\Sigma_{pb}$ by the relationship:
$$
\rho_{pb}=\Sigma_{pb}/(\sqrt{2\pi} H_{pb}) \ .
$$

Regrouping terms we find:
\begin{equation}
\dot{M}_{3D}=\dot{M}_{2D} \left({{\pi r_{eff}}\over{2\sqrt{2\pi}H_{pb}}}\right)\ .
\label{Mdot3D}
\end{equation}

\vskip 18pt
\noindent{\bf Transition from the 2D to the 3D regime}

From the last formula, it is obvious that the transition occurs when:
$$
{{\pi r_{eff}}\over{2\sqrt{2\pi}H_{pb}}} < 1\ .
$$
This sets a condition on $H_{pb}$. The latter is related to the disk scale height $H_g$ by the relationship: 
\begin{equation}
H_{pb}=H_g\sqrt{\alpha/\tau}\ ,
\label{Hpebble}
\end{equation}
(Youdin and Lithwick, 2007), where $\alpha$ is the turbulence parameter of the disk of gas in the  Shakura and Sunyaev (1973) prescription.

{Thus, 2D accretion will occur only for large planets (i.e. large $r_{eff}$) in disks with thin particle layers (i.e. small $H_{pb}$). This is the case if turbulence is weak (i.e. small $\alpha$) and particles large (i.e. large $\tau$). For small planets and thick particle disks, accretion is 3D.}

\vskip 18pt
\noindent{\bf Dependence of the accretion rate on $\tau$}

{We now develop the dependence on the particles' Stokes number $\tau$ of the accretion rates reported in Eq.~(\ref{Mdot2D}) and~(\ref{Mdot3D}). For this purpose, we consider first the most likely case where $v_{rel}=v_{shear}$ (see eq. \ref{vrel}). For the case of 2D accretion, because $r_{eff} \propto \tau^{1/3}$ (eq.~\ref{reff}) and $v_{shear}\propto r_{eff}$ (eq.~\ref{vrel}), formula (\ref{Mdot2D}) gives:
\begin{equation}
\dot{M}_{2D} \propto \tau^{-1/3} \dot{M}_F \ .
\label{m2dtau}
\end{equation}
For the case of 3D accretion, because $H_{pb}\propto \tau^{-1/2}$ (eq~\ref{Hpebble}), formula (\ref{Mdot3D}) gives:
\begin{equation}
\dot{M}_{3D} \propto \tau^{1/2} \dot{M}_F \ .
\label{m3dtau}
\end{equation}
If instead $v_{rel}=\Delta v$ (this happens only when $r_{eff}$ is very small), it is easy to see that the expressions for $\dot{M}_{2D}$ and $\dot{M}_{3D}$ become proportional to $\tau^{-2/3}$ and $\tau^{1/6}$, respectively. 

Thus, it is important to realize that in 3D accretion, the smaller is the value of $\tau$ (i.e. smaller particles or higher gas density) the less efficient is the accretion. In 2D accretion, it is the opposite. }

\vskip 15pt
With these formul\ae\ in hand, we can now design a scenario which may explain the great Solar System dichotomy and test it with simulations. 

\subsection{Model scenario and simulations}
\label{sims}

Planetary embryos and cores grow from the pebbles that flow through their orbits. The basic assumption of the model we propose is that the pebble flux is quite different within and beyond the snowline (respectively for $r<r_{\rm ice}$ and $r>r_{\rm ice}$ where $r_{\rm ice}$ is the location of the snowline). Beyond the snowline there is a flow of icy pebbles. Lambrechts and Johansen (2014) have developed a pebble coagulation and drift model based on the work of Birnstiel et al. (2012). They found that, at $t=10^5$--$10^6$~y of the disk's evolution, icy pebbles at $\sim 4$~AU are a few cm in radius with $\tau\sim 10^{-1.5}$. They carry a mass flux of $\dot{M}_F\sim  10^{-4} M_E/y$. Here we will assume that the snowline is at 3.5~AU, consistent with the Grand Tack model of evolution of the giant planets and formation of the terrestrial planets (Walsh et al., 2011; Jacobson and Morbidelli, 2014), but the exact location of this line does not have practical importance in what follows. 

When the pebbles come to the snowline, it is unlikely that they stop drifting towards the Sun. The drift of particles can stop only in presence of a pressure trap, where the radial pressure gradient of the gas is positive, so that the gas is locally super-Keplerian. It has been proposed (Kretke and Lin, 2007) that the snowline can create a pressure bump because the condensation of icy grain could strongly reduce the viscosity of the gas, thus increasing the gas surface density at the snowline location. However, improved 3D simulations of the disk structure in presence of viscosity transitions (Bitsch et al., 2014b) show that this is true only in presence of extreme viscosity transitions, with ${{\rm d}\log(\nu)\over{{\rm d}\log(r)}}< -28$ where $\nu$ is the viscosity. It is very unlikely that the condensation of icy grains can have such a dramatic effect on the viscosity.  

Thus, we assume that the icy pebbles drift past the snowline. When they penetrate into the warm part of the disk, the ice sublimates away. This decreases the muss flux by a factor of $\sim 2$ (Lodders, 2003). Moreover, we assume that the icy pebbles are a mixture of ice and numerous silicate grains, hold together by the ice; thus, by sublimation, the icy pebbles release silicate grains which are much smaller than the original icy pebbles; these grains have a significantly smaller Stokes number and are much more coupled to the gas. 

{It is unclear a priori what the typical size of the silicate grains could be. However, observations can give us some hints. Ordinary chondrites are made mostly of chondrules, which are mm-sized particles (Friedrich et al., 2014). We do not know whether the sublimation of the icy pebbles released chondrules, or just the precursor grains of chondrules, but in terms of size the two are probably very similar. Moreover, coagulation experiments show the effectiveness of a mm-bouncing barrier for silicate particles (G{\"u}ttler, 2009), so it is likely that silicate particles could grow to mm-size relatively easily but not beyond (unlike icy particles which can easily grow to larger sizes; Wada et al., 2009; Ros and Johansen, 2013). For all these reasons, we assume that the silicate particles released by the sublimation of the icy pebbles are millimeter in size.}  These particles would have a Stokes number $\tau\sim 10^{-2.5}$ at 3.5~AU. For simplicity we assume a unique size (i.e. value of $\tau$) for the icy pebbles and the silicate grains, rather than a distribution peaked around the considered values. 

We consider the mass growths of two embryos on either side of the snowline, at 3.5 AU\footnote{ Obviously it makes no sense that both bodies are at 3.5 AU. We do this assumption to suppress all dependencies on heliocentric distance and highlight only the consequences of the snowline effects on pebble size and mass flux.}. Both bodies have initially the same mass, which we set to half lunar masses (0.005$M_\oplus$),or about to 2.5 Pluto masses. This choice of the initial mass is rather arbitrary, but motivated by a few considerations: (a) we wish the masses of the two bodies to be the same, so that the observed differences in the growth rate are due solely to the snowline effects and not to a difference in the initial masses; (b) we wish a mass significantly smaller than the target masses for the embryo and the core of the giant planet (0.1 and $\sim 10 M_\oplus$, respectively); (c) but at the same time we wish the bodies to be massive enough to be from the beginning in the Hill regime, to avoid the complication of the transition from the Bondi regime (for the typical value of $\eta=0.0027$ that we assume here, the transition between the two regimes occurs at $M=0.0037M_\oplus$). An additional motivation will be given in sect.~\ref{multiple}. {Modeling the formation of these bodies is beyond the scope of this paper. However, a recent work (Johansen et al., 2015) showed that they can be produced by a combination of streaming instability and pebble accretion. Our paper focuses on how subsequent pebble accretion, with different mass fluxes and pebble sizes, can lead to the great Solar System dichotomy between giant planet cores and terrestrial planet embryos.} 

{Because the dependence of the accretion rate on $\tau$ is different in the 2D and 3D accretion regimes, we consider separately two cases: first that of a turbulent disk, for which the particle layer is thick and most of the accretion occurs in 3D and then that of a low-turbulent disk, for which the particle layer is thin and accretion is 2D from the beginning.
}

\subsection{Turbulent disk}

\begin{figure}[t!]
\centerline{\includegraphics[height=8.5cm]{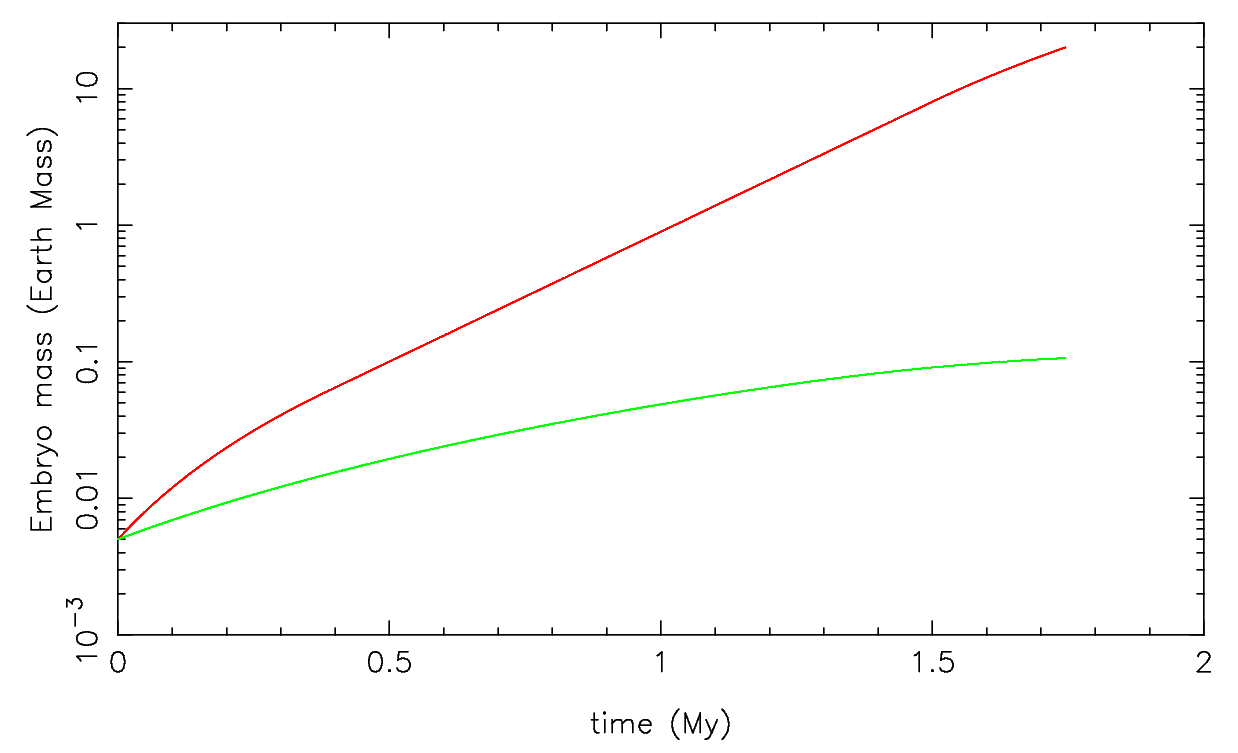}}
\caption{\small Mass growth due to a flux of pebbles  of two embryos located on opposite sides of the snowline, here assumed to be at 3.5~AU. The parameter $\eta$ is assumed to be $0.0027$. The red curve is for an embryo beyond the snowline, with a flux of $2\times 10^{-4} M_E/y$ in icy pebbles with $\tau=10^{-1.5}$. The green curve is for the inner embryo, for which the mass flux is reduced to $1/2(1-F)$ relative to the outer embryo, and the pebbles' Stokes number is $\tau=10^{-2.5}$. Here $F$ is the fraction of the pebble flux which is intercepted by the outer embryo.  When the outer embryo reaches 20$M_\oplus$ the flux of pebbles towards the inner embryo stops (Lambrechts et al., 2014).  Thus, the inner embryo stops growing, and the simulation is stopped at this point. According to Lambrechts et al., 2014), the outer embryo then starts accreting gas efficiently, to become eventually a giant planet.}
\label{equalembryos}
\end{figure}

Embryos with $M=0.005 M_\oplus$ at 3.5 AU have a Hill radius $R_{H}=0.006$~AU. We assume an aspect ratio for the disk of gas of 0.05 and a turbulence parameter $\alpha=10^{-3}$. Thus, the scale-height of the disk of icy pebbles is 0.03 AU (see eq. \ref{Hpebble}) and that of the silicate grains is $\sim 3$ times larger. So, both embryos start growing in the 3D regime. 

{The accretion history of the two embryos is shown in Fig.~\ref{equalembryos}. The accretion of the inner embryo is penalized relative to that of the outer body for three reasons. First, the inner body sees roughly mm-sized particles, which correspond to a value of $\tau$ 10 times smaller than of the pebbles beyond the snowline. Remember that in 3D accretion $\dot{M}_{3D} \propto \tau^{\beta}$, with $\beta=1/6$ or $1/2$ depending on the the scaling of $v_{rel}$; so a smaller $\tau$ implies a smaller accretion rate. Second, the total mass flux of particles is assumed to be two times smaller due to ice sublimation. Third, the flux of pebbles inside the orbit of the outer embryo is $(1-F)$ times that going towards the orbit of the outer embryo, where $F$ is the filtering factor (Morbidelli and Nesvorny, 2012), namely the fraction of the pebble flux intercepted by the outer embryo, which is computed numerically from the embryo's accretion rate.}

{Thus, it is no surprise to see in Fig.~\ref{equalembryos} that the outer embryo grows much faster than the inner one. The ratio between the initial growth rates is $\sim 3$. During their accretion histories,} the outer embryo passes into the 2D accretion regime at 1.5~My. The inner one remains in the 3D regime all the time. The simulation is stopped when the outer embryo reaches 20$M_\oplus$. {This is because embryos this massive can open a partial gap in the gas distribution, enough to reverse the pressure gradient at its outer border. This reversal in the pressure gradient makes the disk locally super-Keplerian and the inward radial drift of pebbles stops there (Lambrechts et al., 2014).} Consequently, the inward embryo has to stop accreting, because the pebble flux is interrupted. Here, the outer embryo reaches 20$M_\oplus$ at $t=1.75$~My, but this time can be changed by changing the pebble mass flux (here assumed to be constant over time for simplicity and equal to $2\times 10^{-4} M_E/y$). The time scales linearly with the inverse of the assumed mass flux. Thus, what is important is not the time at which the outer embryo reaches 20$M_\oplus$ but the mass that the inner embryo has when this happens. As one can see, the final mass of the inner embryo (green) is of the order of a Mars mass. Thus, we have reproduced the great dichotomy of the Solar System. 

We have tested different simulation set-ups in order to understand the role played by each of our assumptions.  Without reducing the value of $\tau$ by an order of magnitude for the pebbles inside of the snowline, the inner embryo would have grown 3.5 times more massive. Without reducing the overall pebble mass flux by a factor of 2 at the snowline, the inner embryo would have grown five times more massive (the final mass does not scale linearly with the mass flux because a more massive object accretes more efficiently). Without both these reductions (the only difference between the inner and outer embryos now being just the filtering factor $1-F$), the inner embryo would have grown 100 times more massive (i.e. would have reached 10~$M_\oplus$). Restoring the nominal set-up concerning the pebble flux, but reducing the initial masses of the embryos by one order of magnitude\footnote{ Notice that these bodies initially would accrete in the Bondi regime, i.e. $R_{GP}=R_B$. Here, we still apply formula (\ref{reff}) even if accurate only for the Hill regime.}, we find that the outer embryo would have reached 20$~M_\oplus$ in 2.35~My, but the inner embryo would have grown only to $1.5\times 10^{-2} M_\oplus$. If the initial mass of the embryos is this small, bringing the inner embryo to Mars-mass requires no reduction in $\tau$ inside of the snowline. 

\subsection{Quiescent disk}
 
{Let's now consider a disk with a very low turbulence, with $\alpha=10^{-6}$. In this case, the height of the icy particle layer at 3.5 AU is 0.001 AU (see eq. \ref{Hpebble}) and that of the silicate grains is $\sim 3$ times larger. In this case accretion is 2D for both the embryos from the beginning of the simulation. In this case, the inner embryo has a comparative advantage relative to the outer one, because $\dot{M}_{2D} \propto \tau^{\beta}$ with $\beta=-1/3$ or $-2/3$ depending on the scaling of $v_{rel}$; thus smaller particles result is faster accretion. There is nevertheless a factor of 2 reduction in the total mass flow of solids, due to ice sublimation. The filtering factor $(1-F)$ is negligible at the beginning of the simulation. 

\begin{figure}[t!]
\centerline{\includegraphics[height=8.5cm]{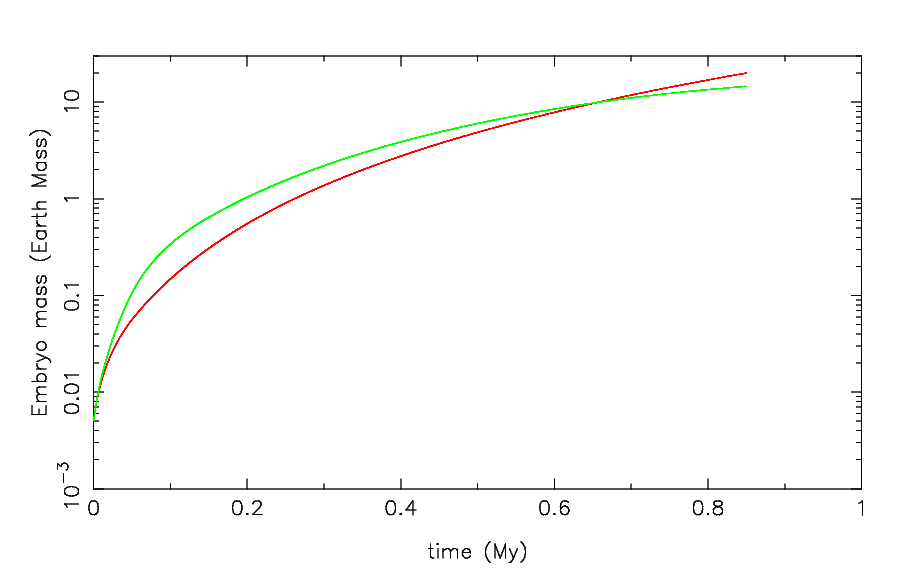}}
\caption{\small The same as Fig.~\ref{equalembryos} but for a much thinner particle disk. Here the height of the layer of icy particles at 3.5~AU is 0.001~AU, and that of silicate particles is $\sim 3$ times as much. This corresponds to a very low turbulence in the disk, with $\alpha=10^{-6}$. The result fails to explain the great dichotomy of the solar system. }
\label{noturb}
\end{figure}

Fig.~\ref{noturb} shows the result in this case. The inner embryo starts to accrete slightly faster than the outer one. It is only later, when the outer embryo exceeds the mass of Mars, that the factor $(1-F)$ starts to become non-negligible; this reduces the growth rate of the inner embryo and the mass ratio between the outer and the inner embryos starts to increase. At the end of the simulation, when the outer embryo reaches $20 M_\oplus$, both embryos are of comparable mass. Clearly, this case does not reproduce the great dichotomy of the Solar System. 

Therefore, our explanation is valid only if the proto-solar disk was not so quiet. A value of $\alpha=10^{-6}$ is very small, even for a so-called {\it dead zone} where the Magneto-Rotational Instability is not active (Stone et al., 2000). In fact, some turbulence can be sustained in the dead zone due to a number of effects, e.g. diffusion of the magnetic accretion stress (Turner et al., 2007),  baroclynic instability (Klahr and Bodenheimer, 2003), vertical shear instability (Nelson et al., 2013). In particular, Stoll and Kley (2014) estimated that the vertical shear instability can sustain turbulence with a strenght equivalent to a parameter $\alpha$ of about few times $10^{-4}$. 

Setting $\alpha=10^{-4}$, we find that the inner embryo grows only to twice the mass of Mars, with an accretion history starting (and remaining) in the 3D regime. For $\alpha$ smaller than this threshold, a significant part of the accretion history of the inner embryo becomes 2D and a clear difference in mass relative to the embryo beyond the snowline disappears, as in Fi.g~\ref{noturb}. Thus, in the rest of this paper we will assume $\alpha=10^{-3}$, with the understanding that the results are quite similar for $\alpha \gtrsim 10^{-4}$.}

\subsection{Multiple embryos in the inner Solar System}
\label{multiple}

The successful simulations of terrestrial planet accretion (e.g. O'brien et al., 2006; Walsh et al., 2011; Jacobson and Morbidelli, 2014) require that all embryos in the inner solar system had a mass of the order of the mass of Mars, not just the one that was the closest to the snowline. Thus, in this section we examine the radial dependence of the mass growth of a chain of embryos located from 0.7 AU up to the snowline location at 3.5 AU. 

Here, there are again a number of assumptions to make. We assume that initially there are 40 embryo ``seeds'' with a mass of 0.005~$M_\oplus$, equally spaced between the two distance limits reported above. { We chose the number 40 because this is approximately the number of embryos used in simulations of terrestrial planet formation (e.g. Jacobson and Morbidelli, 2014). We chose the mass of 0.005~$M_\oplus$ because it is generally accepted that the inner solar system contained about 1,000 Ceres size bodies (Wetherill, 1992) with a mass $\sim 2\times 10^{-4} M_\oplus$ so that, assuming a cumulative size distribution $N(>D)\propto 1/D^3$ like that of the largest main belt asteroids, we expect 40 bodies with masses $\ge 0.005 M_\oplus$. Obviously we do not expect that all embryo seeds had exactly the same mass, but in this experiment it is important that all masses are equal at the beginning of the simulation so that the observed differences in growth rates are due solely to pebble accretion effects and not to initial mass differences.} Moreover, as the silicate grains drift towards the Sun we assume that they do not evolve in size by coagulation or fragmentation. This is because there is no clear evidence for the dependence of the chondrule's size with the heliocentric distance of the chondritic parent bodies (Friedrich et al., 2014). However, because the density of gas changes, the Stokes number decreases as the particles decrease their heliocentric distance. Given the shallow radial density profile of the gas defined in section~2 we assume that the Stokes number of a particle is $\tau\propto \sqrt{r}$. 

\begin{figure}[t!]
\centerline{\includegraphics[height=8.cm]{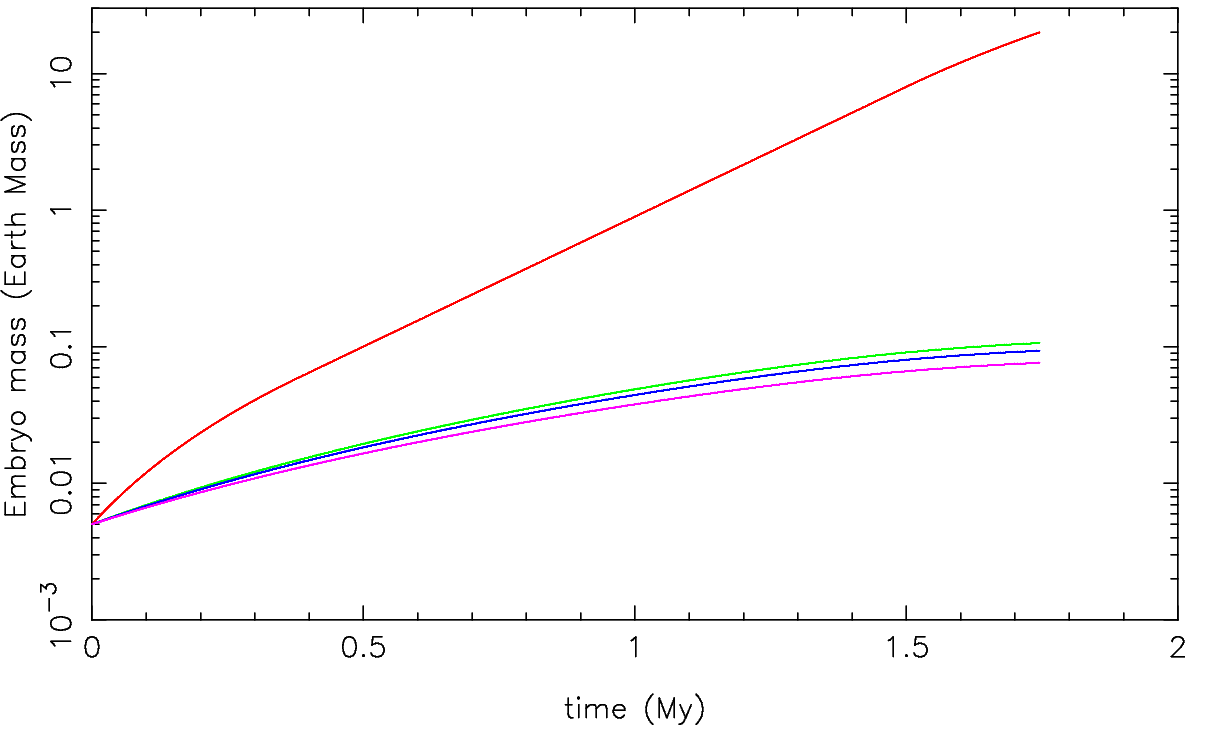}}
\centerline{\includegraphics[height=8.cm]{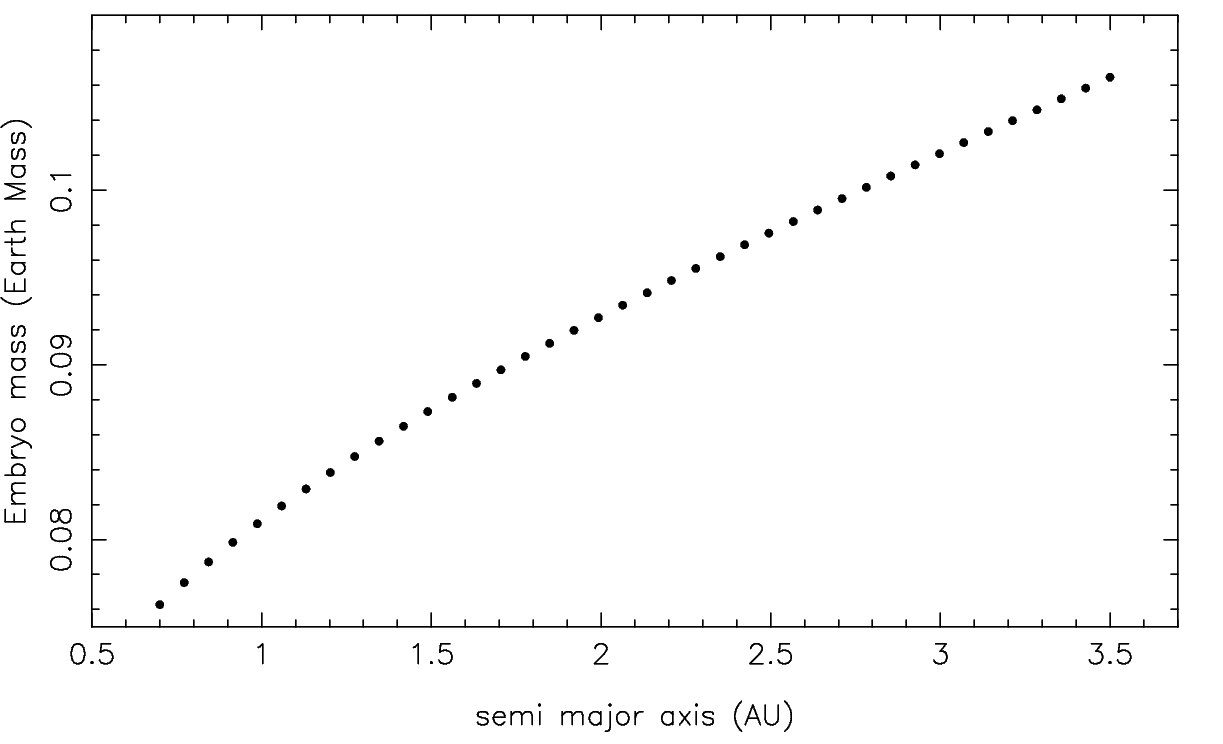}}
\caption{\small Top: the same as Fig.~\ref{equalembryos} but now showing the evolution of 3 embryos in a chain of 40. The red curve is for the giant planet core and the green curve is for the embryo \#40 (the outermost one at 3.5 AU), both already shown in Fig.~\ref{equalembryos}. The blue curve is for the embryo \#20 at 2.1 AU, and the magenta curve is for the embryo \#1 at 0.7 AU. Bottom: the final mass distribution of the embryos as a function of heliocentric distance.}
\label{40embryos-noeta}
\end{figure} 
 
In a first step, in order to understand the role of each parameter, we assume that $\eta$ in (\ref{eta}) is independent of radius ($\eta=0.0027$ as in the previous simulations). From formula (\ref{eta}), one can see that the parameter $\eta$ would indeed be constant if a disk had $H_g\propto r$ and $P$ were a power-law function of $r$. The results is shown in Fig.~\ref{40embryos-noeta}, which reports in red the growth of the giant planet core beyond the snowline (same as in the previous figures), in green the growth of the outermost embryo at 3.5~AU (same as in Fig.~\ref{equalembryos}), in blue the embryo in the middle of the chain (at 2.1 AU) and in magenta the innermost embryo (at 0.7 AU). As one can see, the growth rate decreases {slightly} approaching the Sun. This is due to the decrease in $\tau$, but also because each embryo receives a flux of pebbles reduced by the other embryos located upstream relative to its position. In other words, counting the embryos from the innermost one, the pebble flux across the orbit of embryo  \#$i$ is $F_i=\prod_{j=i+1}^{j=40} (1-F_j)$, where $F_j$ is the filtering factor of embryo \#$j$.

\begin{figure}[t!]
\centerline{\includegraphics[height=7.5cm]{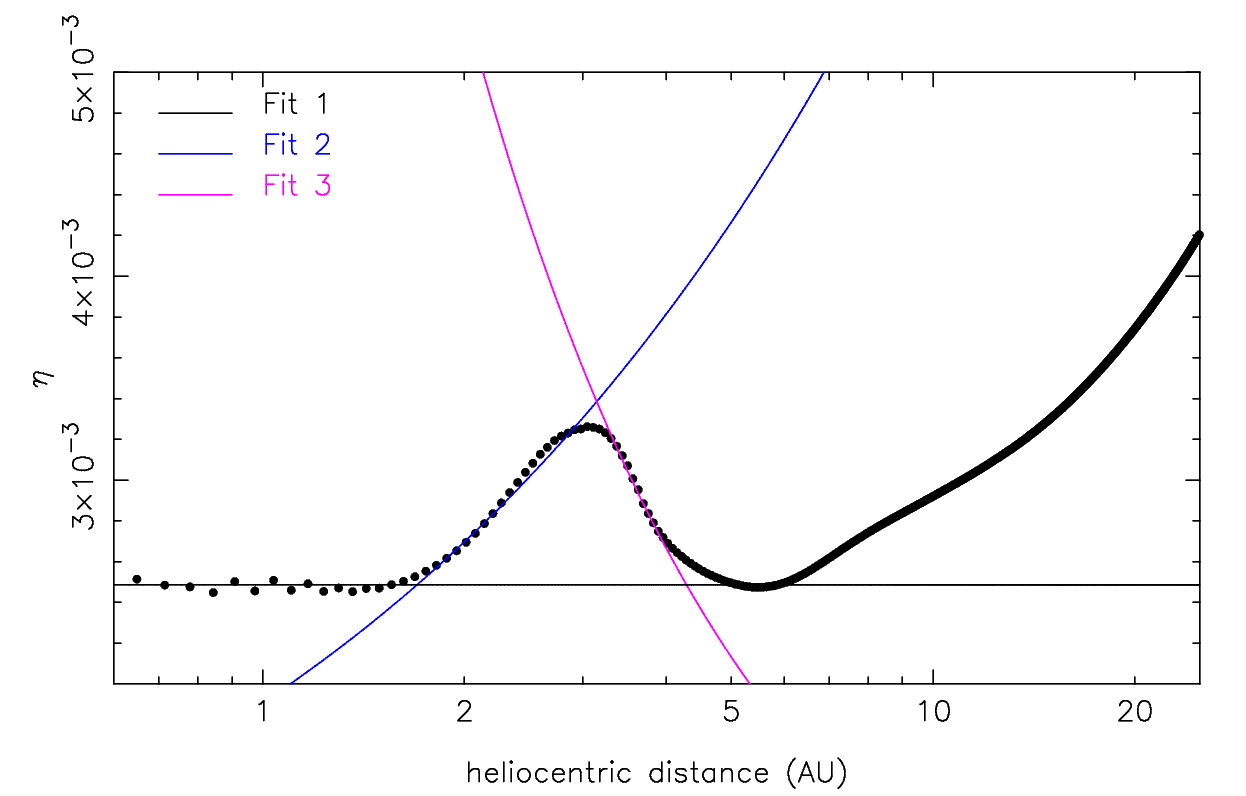}}
\centerline{\includegraphics[height=7.5cm]{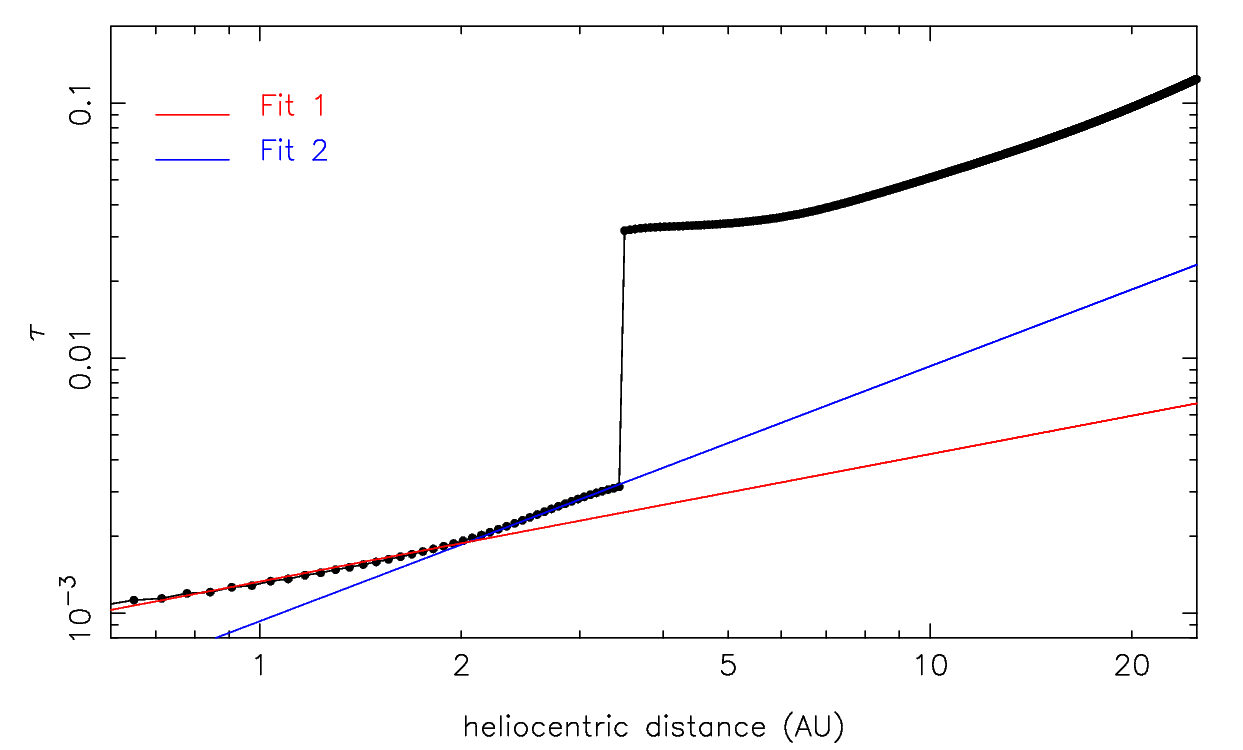}}
\caption{\small Top: the dots show the radial dependence of $\eta$ according to Bitsch et al. (2015) for a disk with $\dot{M}=3.5\times 10^{-8} M_\odot /$y. The colored solid lines represent analytic fits valid over specific intervals of heliocentric distance. Fit 1: $\eta=0.002487$; Fit 2: $\eta=0.002487\sqrt{r/1.7}$; Fit 3: $\eta=0.00328/(r/3.25)$. %; Fit 4: $\eta=0.00245 (r/5.5)^{0.3}$; Fit 5: $\eta=0.0038 \sqrt{r/20.5}$. 
Bottom: the same but for the radial dependence of $\tau$. Here we assume a jump in pebble size by a factor of 10 at 3.5 AU, as illustrated by the black curve connecting the dots. Here the fits for $r<3.5$~AU are: Fit 1: $\tau=0.00133\sqrt{r}$; Fit 2: $\tau=0.00093\,r$.}
\label{etaBitsch}
\end{figure} 

However, in a realistic disk, $\eta$ is not fully independent of $r$, because the aspect ratio of the disk of gas $H_g/r$ has oscillations as a consequence of the opacity changing with temperature (Bitsch et al., 2014a, 2015). The same is true for the radial dependence of $\tau$. Fig.~\ref{etaBitsch} shows theses dependence for the disk studied in Bitsch et al. (2015) corresponding to an accretion rate on the star of $3.5\times 10^{-8} M_\odot/$y. We chose this disk among those studied in that work because it has the snowline at 3.5 AU, as we assumed above. The figure also reports various analytic fits, valid in specific ranges of heliocentric distances, that we will use below.

\begin{figure}[t!]
\centerline{\includegraphics[height=8.cm]{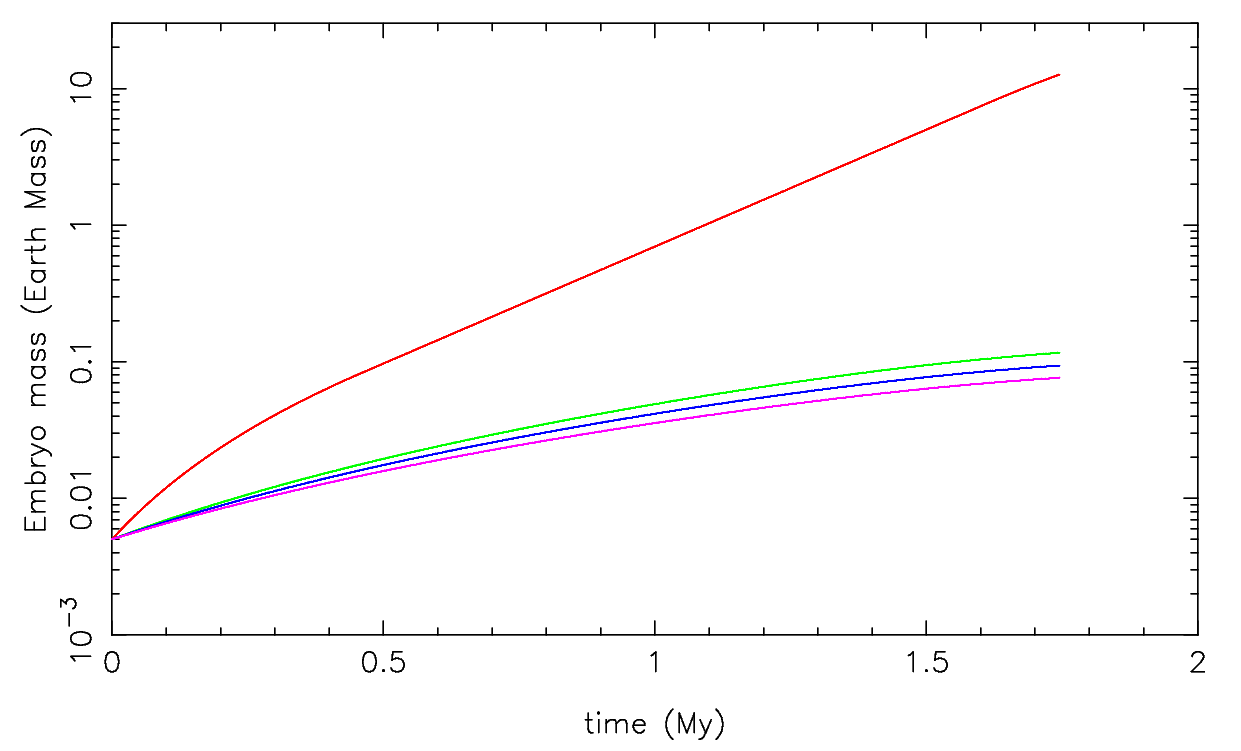}}
\centerline{\includegraphics[height=8.cm]{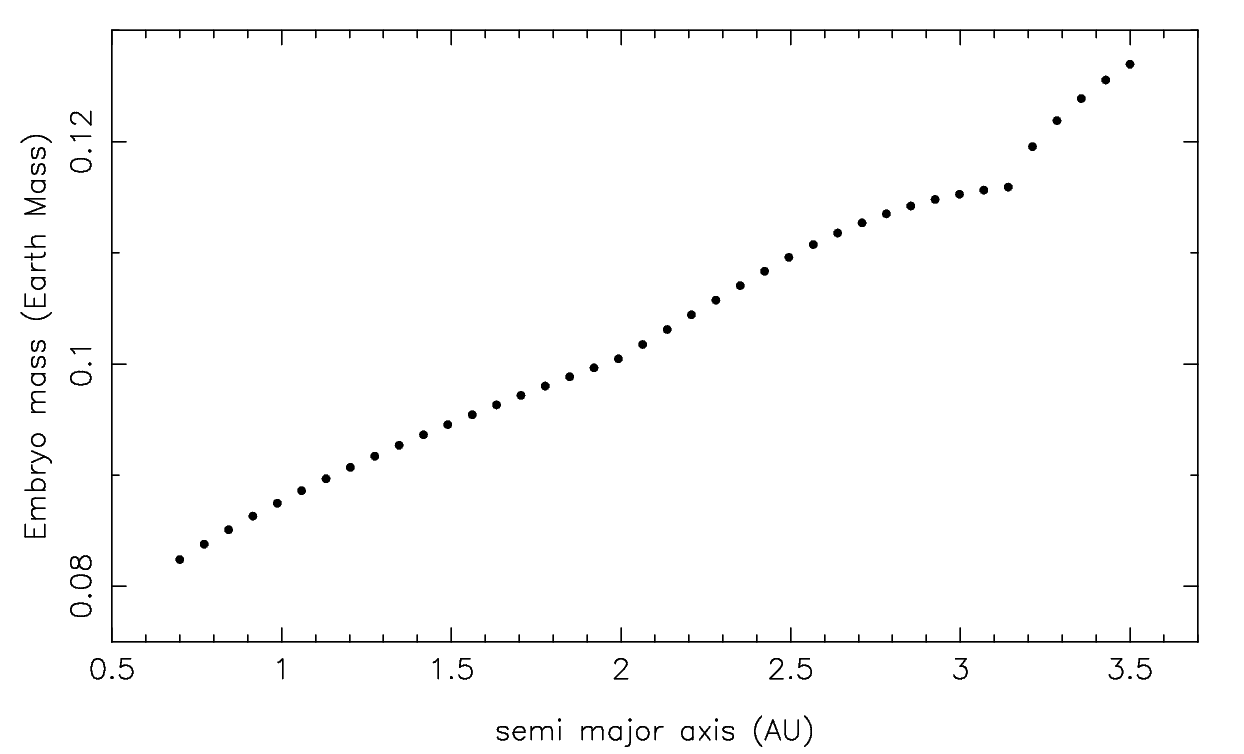}}
\caption{\small The same as Fig.~\ref{40embryos-noeta} but now assuming $\eta(r)$ and $\tau(r)$ as given by the fits shown in Fig.~\ref{etaBitsch}.}
\label{40embryos-eta}
\end{figure} 
 
For $\eta$ we assume Fit 1 for $r<1.7$~AU, Fit 2 for $1.7<r<3.2$~AU and Fit 3 for $3.2<r<4$~AU.  For $\tau$ we assume Fit 1 for $r<2$~AU and Fit 2 for $2<r<3.5$ AU. The value of $\tau$ for the core placed just beyond the snowline remains equal to $10^{-1.5}$. With this set-up, the results are shown in Fig.~\ref{40embryos-eta}.  The results are quite similar to those shown in Fig.~\ref{40embryos-noeta}. The mass distribution of the embryos has a clear kink at 3.15 AU, where the $\eta$ function has a local maximum.  In fact, between 3.15 and 3.5 AU, $\eta$ sharply decays with increasing heliocentric distance $r$. Pebble accretion is strongly favored if $\eta$ is small. Thus, the mass distribution of embryos increases sharply with $r$. Instead, for $r<3.15$~AU, $\eta$ increases with increasing $r$. In principle, this would give a mass distribution of embryos decreasing with increasing $r$. But remember that, as the outer embryos start to grow, their filtering factors $F_j$ grow as well and the inner embryos receive a progressively reduced flux of pebbles. Thus, the final mass distribution is still a {moderately} growing function of $r$, but not as steep as beyond 3.15~AU. 

%A growing radial mass distribution of embryos has been advocated by Jacobson and Morbidelli (2014) as a potential solution of the problem of the small mass of Mercury. Further investigations are needed to verify whether a mass distribution of embryos like that of Fig.~\ref{40embryos-eta}b leads to a better reconstruction of the terrestrial planets system than the usually assumed flat distribution. 

\subsection{Caveats and Issues}

We discuss here some aspects of the problem which have been neglected so far: disk evolution, the effect of self-excitation of the orbits of the embryos, the effect of radial migration and the mass distribution of the giant planet cores.

\subsubsection{Disk evolution}

{ Over time the disk evolves and loses mass. The stellar accretion rate decreases (Hartmann et al., 1998). Consequently, viscous heating drops and the transition between the inner region of the disk with roughly constant aspect ratio and the outer flared region dominated by stellar irradiation moves towards the star (Bitsch et al., 2015). This effect changes the dependence of $\eta$ as a function of radius. The snowline location also moves towards the star. Moreover, with decreasing gas density, $\tau$ increases for a given pebble size but on the other hand only smaller pebbles may survive in the disk (Lambrechts and Johansen, 2014). 

In this paper we have neglected all of this. The goal here was to give a proof of principle that pebble accretion can explain the dichotomy in the mass distribution of embryos and cores within and beyond the snowline and for this reason we preferred to remain simple, working in the framework of a snapshot of the disk's structure, taken more or less in the middle of its evolution. A more quantitative modeling of planet accretion accounting for disk's evolution remains to be done and is left for a future work.}

\subsubsection{Embryos' self-excitation}
\label{self}

Kretke and Levison (2014b) pointed out that if the embryos acquire orbits with significant eccentricities or inclinations they stop accreting pebbles. In fact, the relative velocities between the embryos and the pebbles become large, which drastically decreases the Bondi radius (\ref{Bondi}). They proposed that this process is the key to explain why only a handful of giant planets cores formed. In their simulations a system of multiple proto-cores, while growing, starts to self-excite their orbital distribution. Because of dynamical friction, the most massive proto-cores remain on the most circular and co-planar orbits, while the less massive ones acquire larger eccentricities and inclinations. This shuts down the accretion of the latter, allowing only a few cores to grow to large masses. 

Given the population of multiple embryos that we consider here in the inner solar system, it is necessary to verify whether their mutual self-excitation of the orbits could change drastically their accretion history. The simulations presented above are not dynamical simulations. They just implement the analytic formul\ae\ described in Sect.~\ref{formulae}. Thus, we address the issue of orbital self-excitation with a separate N-body simulation. 

We assume a system of 40 Mars-mass embryos, from 0.7 to 3.5~AU, thus separated by 6.75 mutual Hill radii (the mutual Hill radius being defined as $(a+a')/2 [(M+M')/(3M_\odot)]^{1/3}$ where $a$, $M$, and $a'$, $M'$ are the semi major axes and masses of two adjacent embryos). Their initial orbits are circular, inclined by $10^{-4}$ radians and with random orientations (to avoid that the problem is planar). We don't include Jupiter in this simulation because at the stage we are interested in Jupiter was just a core of less than 20 Earth masses.  

We simulate the evolution of this system of embryos for 2~My with a version of the {\it Symba} integrator (Duncan et al., 1998) modified to account for the tidal damping of eccentricities and inclinations exerted by the embryos' gravitational interaction with the disk of gas. For massive bodies like the embryos, this is the most important damping effect, whereas gas drag is negligible (the opposite is true for planetesimals). The tidal damping formul\ae\ are those issued in Cresswell et al. (2007) and Cresswell and Nelson (2008) from hydrodynamical simulations. For this simulation, we assume that disk of gas has a density of 2400g/cm$^3$ at 1 AU, corresponding to the MMSN, but with a radial profile of $1/\sqrt{r}$ expected for an accretion disk (see sect.~\ref{disk}). We neglect the disk torque inducing radial migration (migration being discussed in sect.~\ref{migration}), although there is some radial migration associated with the eccentricity damping (for instance the innermost embryo is at the end of the simulation at 0.69~AU and the outermost one at 3.36~AU). 

The system of embryos remains on stable orbits. No merger events have been recorded. At the end of the simulation the embryo with the largest eccentricity has $e=0.0018$. The median value of the eccentricities of the embryos is 0.0006. All inclinations are below $10^{-4}$ degrees. Thus, the orbital excitation in terms of deviation from the Keplerian velocity ($e V_K$) is smaller (in most cases much smaller) than the differential velocity of the gas relative to the embryos ($\eta V_K$),  with $\eta$ given in Fig.~\ref{etaBitsch}a. Keep in mind that this simulation likely overestimates the real orbital excitation of the embryos. In fact, during most of their growth, the embryos are less massive than one Mars-mass and, even at the end of the simulation illustrated in Fig.~\ref{40embryos-eta}, not all embryos have reached this value. Thus we conclude that self-excitation of the embryo's orbits is not an important issue affecting their growth. Self-excitation would have become important if the embryos had grown well beyond the mass of Mars, as it is the case beyond the snowline. Thus our results do not contradict those of Kretke and Levison (2014b). It is because growth by pebble accretion is strongly reduced within the snowline that the inner solar system ended up with numerous small-mass embryos, while the outer solar system, where pebble-accretion is more vigorous, produced large cores but only in small number.

\subsubsection{Migration}
\label{migration}

In all the calculations presented in this paper, as well as in the N-body simulation, we have neglected planet migration. Planet migration is due to the tidal interaction with the disk of gas. In first approximation, the migration speed is proportional to the planet's mass (Tanaka and Ward, 2002). For the gas disk considered to build Fig.~\ref{etaBitsch}, a Mars-mass object would migrate from 1.8AU down to 1.3AU in $\sim 3$ Myr. A migration range of 0.5~AU looks large compared to the typical embryo-embryo distance of $\sim 0.1$~AU. However, remember that: (i) all embryos would migrate at a comparable rate, so their mutual distances would change much more slowly and (ii) embryos are much smaller than Mars-mass for most of the lifetime of the disk, thus they would migrate much less then reported above. In conclusion, we think we can safely neglect their migration during the growth process at the stage of a proof of concept paper as this one.

The situation for the giant planet core is different. According to the classical Type-I migration model (Goldreich and Tremaine, 1980; Tanaka and Ward, 2002) the core should migrate into the inner solar system  very quickly. But a new result by Benitez-Llambay et al. (2015) shows that a fast-accreting body receives a positive torque from the disk, due to the fact that it is very hot and it heats the surrounding gas in an asymmetric way (given that the disk is sub-Keplerian). Thus it is possible that the core refrains from migrating inwards. Once the core becomes more massive than a few Earth masses, this ``accretional torque'' should lose relative importance. But at this stage, the thermal structure of the disk becomes relevant, as it can excite an entropy-driven corotation torque on the core (Paardekooper and Mellema, 2006; Paardekooper et al., 2010, 2011; Masset and Casoli, 2009, 2010). Bitsch et al. (2014a, 2015) showed that the core should position itself a few AUs beyond the snowline (approximately where $\eta$ reaches the minimum value; see Fig.~\ref{etaBitsch}a), at a location where all the different torques that it suffers from the disk cancel each other out. 

As the disk evolves and cools, the  mass range of cores for which the torques cancel out  shrinks  (Bitsch et al., 2015) so that, eventually, a core is released to Type-I migration. Thus, the assumption that the core grows to 20$M_\oplus$ without migrating is, by no doubt, simplistic. We have done it because migration is complicated (and not fully understood, hence the appearance of new effects as the ``accretional torque'') and we wanted to keep the model simple, as a proof of concept. The issue of core migration will be the object of a future work. Nevertheless, we can not resist to make a few speculations about the effect of migration at the very end of this paper. 

\subsubsection{Cores mass distribution}

In the present paper we have considered only one core beyond the snowline. But at least 4 cores formed in the outer solar system. For the goal of investigating the mass contrast between the embryos within the snowline and the first core outside of the snowline it was enough to fix the mass flow of pebbles across the orbit of the considered core, regardless of how that flow had been reduced by the presence of additional cores further out. We stress that the mass flow is related to the time at which a certain mass is achieved by a given body. Thus, reducing the mass flow over time (for instance because the cores outside of Jupiter intercept an increasing fraction of the overall flow) is just equivalent to stretching the x-axis of Fig.~\ref{40embryos-noeta} and~\ref{40embryos-eta} in a non linear way. The masses of the inner solar system embryos as a function of the mass of the core of Jupiter would not change. 

However, one might wonder what would be the relative mass growth among the cores beyond the snowline. As we have seen, in the inner solar system the accretion rate of embryos grows with heliocentric distance. But the architecture of the Solar System, with Jupiter having accreted a more massive atmosphere than Saturn and Saturn having accreted more gas than Uranus and Neptune, suggests that the core of Jupiter grew the fastest, then Saturn's, then those of the ice-giant planets. The growth of these cores has been already modeled in Lambrechts and Johansen (2014). They found indeed that the innermost core grows the fastest. The reason for this is twofold. The most important factor is that beyond the snowline the disk is flared, so that $\eta$ grows rapidly with distance (see Fig.~\ref{etaBitsch}a). The second factor is that icy pebbles grow as they drift, so that their value of $\tau$ increases with decreasing distance (see Lambrechts and Johansen, 2004, Fig. 2) {which, in a moderately turbulent disk in which most of the accretion proceeds in a 3D mode, leads to more efficient accretion}. Instead, within the snowline $\eta$ is almost constant and $\tau$ decreases with decreasing $r$. Our calculations, assuming the $\eta$-function illustrated in Fig.~\ref{etaBitsch}a and the $\tau$-function of Lambrechts and Johansen (2014) indeed confirm that, beyond 5~AU cores formed by pebble accretion have a growth rate decreasing with increasing heliocentric distance. 
%Kretke and Levison (2014b) also found a similar mass ranking in their simulations of the pebble-accretion process, starting with a large number of proto-cores and allowing for the self-stirring of the population.   

\section{Conclusions}

In this paper we have investigated which mode of planet growth can best explain the {\it great dichotomy of the Solar System}. This dichotomy stems from the realization that, in order to form the giant planets, solid cores of approximately 10--20 Earth masses had to form in the outer solar system within the lifetime of the gaseous protoplanetary disk, whereas in the inner Solar System planetary embryos had a mass only of the order of that of Mars (as deduced from the short timescale of Mars accretion and from the mass of the embryo that was involved in the Moon-forming collision with our planet). How to get a ratio of $\sim 100$ between the masses of the protoplanets in the outer and the inner solar system respectively is not obvious. 

We have first considered the classic mode of protoplanet growth, that is ordered accretion of planetesimals aided by the so-called gravitational focusing factor. This leads to the process of runaway growth (Greenberg et al., 1978), followed by that of oligarchic growth (Kokubo and Ida, 1998). We have shown in Sect.~2 that the runaway growth process is much faster in the inner disk than in the outer disk, in sharp contrast with the great dichotomy. Only if the process of oligarchic growth is brought to completion, with each embryo accreting basically all the solid material in its annulus of influence, one can expect more massive bodies further out. This is unlikely to happen, particularly in the giant planet region (Levison et al., 2010, 2012). Nevertheless we find that, if oligarchic growth had reached completion everywhere, the mass of the embryos would have increased progressively with heliocentric distance. Thus, in order to have a body approaching 10$M_\oplus$ beyond the snowline, Earth-mass embryos should have formed in the asteroid belt. Consequently, the resulting mass distribution in the Solar System would not have looked like anything that has been considered so far in successful models of giant planet or terrestrial planet formation. 

A second process for the formation of giant planet cores and planetary embryos is pebble-accretion (Lambrechts and Johansen, 2012). In that process, the most massive planetesimals accrete small objects (pebble-sized) as the latter drift through their orbits. Accretion of pebbles is very effective due to a combination of gravitational deflection and gas-drag. We have shown that the process of pebble-accretion can explain the great dichotomy, provided that a number of assumptions hold true. We have assumed that icy pebbles are a few cm in size, according to the pebble-growth model of Lambrechts and Johansen (2014). When icy pebbles drift through the snowline the ice sublimates and we have assumed that this causes a drop of a factor of 2 in the total mass flux and the release of a large number of silicate grains whose sizes are approximately 1~mm (i.e. about chondrule-size).   With these assumptions we have been able to show that, starting from bodies of {\it equal masses} everywhere in the protoplanetary disk (specifically, we have assumed a mass of 0.5 Lunar-masses), the embryos within the snowline grow to approximately Mars-mass in the time when the first core beyond the snowline grows to 20$M_\oplus$.  {The major condition for this to happen is that there is enough turbulence in the disk so that the thickness of the layer of silicate grains is larger than the effective accretion radius for Mars-mass bodies. This happens if the parameter $\alpha$ characterizing the turbulence strength in the Shakura and Sunyaev (1973) prescription is $\gtrsim 10^{-4}$. According to Stoll and Kley (2014) this condition is fulfilled due to the vertical shear disk instability.}

If this condition holds, the final mass distribution of the planetary embryos within the snowline moderately grows with heliocentric distance. Interestingly, this is the kind of distribution that reproduces best the final mass distribution of the terrestrial planets, including a small Mercury (Jacobson and Morbidelli, 2014). 

{ We kept the model simple on purpose, as a proof of concept. In particular, we have assumed that all the seeds of the future embryos and cores had initially the same masses; we have neglected the disk's evolution; we have considered only one size for the pebbles in each region of the disk.} Obviously, the results could easily change at a  quantitative level by changing the assumptions. Thus, at this stage, in absence of more precise information on the original distribution of the largest planetesimals throughout the disk and the sizes of pebbles and grains, we cannot be firmly predictive of what kind of protoplanets formed through the Solar System within the lifetime of the disk of gas. Nevertheless, we find interesting and intriguing that the pebble-accretion process, with a set of reasonable values for the parameters of the problem, allowed us to reproduce the great dichotomy of the Solar System, with Mars-mass embryos throughout the inner solar system and giant planet cores of 10--20$M_\oplus$ beyond the snowline. Given that the classical process of planetesimal-accretion did not allow us to come even close to the desired result, we conclude that the pebble-accretion process is more likely to explain correctly the formation of protoplanets throughout the Solar System. 

The processes of pebble growth and drift through a disk, as well as the pebble sublimation and splitting when passing through the snowline, should be universal. Thus we expect that any protoplanetary disk developed a sort of dichotomy between the masses of the protoplanets formed within and beyond the snowline, like the one we see (or, better, deduce) in our Solar System. If this is true, the great diversity observed among planetary systems would have to stem from the subsequent evolutions of these protoplanets. We believe that this is likely. For instance, if the cores beyond the snowline do not become giant planets, they would eventually migrate towards the star as the disk loses mass (Bitsch et al., 2015). In this migration process the cores would scramble, disperse and deplete the system of planetary embryos, preventing the accretion of real terrestrial planets (Izidoro et al., 2014); on the other hand the cores, moving into the inner system, would qualify for ``super-Earths'' or ``hot-Neptunes'' (probably low-density ones given their icy nature and their potential hability to accrete some primitive atmosphere), which are the most abundant planets observed around stars (Mayor et al., 2011). If instead the cores become giant planets, their migration history becomes sensitive to their mutual interactions (Masset and Snellgrove, 2001). If only one core becomes giant planet then it should migrate inwards and this should strongly affect the evolution of the inner system of planetary embryos (Fogg and Nelson, 2005; Raymond et al., 2006b). The key aspect of the Solar System evolution was that Jupiter avoided to migrate inwards of a few AU, probably due to the presence of Saturn (Masset and Snellgrove, 2001; Morbidelli and Crida, 2007; Walsh et al., 2011). Thus, Jupiter and Saturn blocked the way to Uranus and Neptune, preventing these cores to migrate into the inner Solar System as the disk cooled (Izidoro et al, 2015). This way, the giant planets remained in the outer Solar System, while the embryos survived in the inner Solar System, eventually leading to the formation of a few terrestrial planets. This specific evolution preserved the vestige of the original dichotomy of protoplanetary masses, set in the primary pebble-accretion process. 

\acknowledgments 

{We thank A. Johansen for insightful discussions and J. Chambers, S. Ida and K. Kretke for their official or unofficial reviews.}  A.M.. was supported by ANR, project number ANR-13--13-BS05-0003-01  projet MOJO(Modeling the Origin of JOvian planets). S. J. was supported by the European Research Council (ERC) Advanced Grant ACCRETE (contract number 290568). M.L. and B.B. thanks the Knut and Alice Wallenberg Foundation for their financial support.

%\begin{thebibliography}{}

\section{References}

\begin{itemize}

\item[--] Agnor, C.~B., Canup, 
R.~M., Levison, H.~F.\ 1999.\ On the Character and Consequences of Large 
Impacts in the Late Stage of Terrestrial Planet Formation.\ Icarus 142, 
219-237. 

\item[--] Allegre, C.J., Manhes, G., Gopel, C. (2008) The major differentiation of the Earth at $\sim$ 4.45 Ga. Earth and Planetary Science Letters 267, 386-398. 

\item[--] Benitez-Llambay, P., Masset, F., Koenigsberger, G. and Szulagyi, J., 2015. Planet heating as a safety net against the inward migration of planetary cores. Nature, in press.

\item[--] Benz, W., Slattery, W.~L., 
Cameron, A.~G.~W.\ 1988.\ Collisional stripping of Mercury's mantle.\ 
Icarus 74, 516-528. 

\item[--] Birnstiel, T., Klahr, H., Ercolano, B.\ 2012.\ A simple model for the evolution of the dust population in protoplanetary disks.\ Astronomy and Astrophysics 539, AA148. 

\item[--] Bitsch, B., Morbidelli, A., Lega, E., Crida, A.\ 2014a.\ Stellar irradiated discs and implications on migration of embedded planets. II. Accreting-discs.\ Astronomy and Astrophysics 564, AA135. 

\item[--] Bitsch, B., Morbidelli, A., Lega, E., Kretke, K., Crida, A.\ 2014b.\ Stellar irradiated discs and implications on migration of embedded planets. III. Viscosity transitions.\ Astronomy and Astrophysics 570, AA75. 

\item[--] Bitsch, B., Johansen, 
A., Lambrechts, M., Morbidelli, A.\ 2015.\ The structure of protoplanetary 
discs around evolving young stars. In press in Astronomy and Astrophysics; ArXiv e-prints arXiv:1411.3255. 

\item[--] Bromley, B.~C., 
Kenyon, S.~J.\ 2011.\ A New Hybrid N-body-coagulation Code for the 
Formation of Gas Giant Planets.\ The Astrophysical Journal 731, 101. 

\item[--] Canup, R.M., Asphaug, E. 2001. Origin of the Moon in a giant impact near the end of the Earth's formation. Nature 412, 708-712. 

\item[--] Chambers, 
J.~E., Wetherill, G.~W.\ 1998.\ Making the Terrestrial Planets: N-Body 
Integrations of Planetary Embryos in Three Dimensions.\ Icarus 136, 
304-327. 

\item[--] Chambers, J.~E.\ 2001.\ 
Making More Terrestrial Planets.\ Icarus 152, 205-224. 

\item[--] Chambers, J.~E.\ 2014.\ Giant 
planet formation with pebble accretion.\ Icarus 233, 83-100. 

\item[--] Cresswell, P., Dirksen, G., Kley, W., Nelson, R.~P.\ 2007.\ On the evolution of eccentric and inclined protoplanets embedded in protoplanetary disks.\ Astronomy and Astrophysics 473, 329-342. 

\item[--] Cresswell, P., Nelson, R.~P.\ 2008.\ Three-dimensional simulations of multiple protoplanets embedded in a protostellar disc.\ Astronomy and Astrophysics 482, 677-690.

\item[--] Crida, A.\ 2009.\ Minimum Mass 
Solar Nebulae and Planetary Migration.\ The Astrophysical Journal 698, 
606-614. 

\item[--] Cuk, M., Stewart, S.T. (2012) Making the Moon from a Fast-Spinning Earth: A Giant Impact Followed by Resonant Despinning. Science 338, 1047. 

\item[--] Dauphas, N., Pourmand, A. (2011) Hf-W-Th evidence for rapid growth of Mars and its status as a planetary embryo. Nature 473, 489-492. 

\item[--] Duncan, M.~J., Levison, 
H.~F., Lee, M.~H.\ 1998.\ A Multiple Time Step Symplectic Algorithm for 
Integrating Close Encounters.\ The Astronomical Journal 116, 2067-2077. 

\item[--] Fogg, M.~J., Nelson, R.~P.\ 2005.\ Oligarchic and giant impact growth of terrestrial planets in the presence of gas giant planet migration.\ Astronomy and Astrophysics 441, 791-806. 

\item[--]  Friedrich, J.~M., 
Weisberg, M.~K., Ebel, D.~S., Biltz, A.~E., Corbett, B.~M., Iotzov, I.~V., 
Khan, W.~S., Wolman, M.~D.\ 2014.\ Chondrule size and related physical 
properties: a compilation and evaluation of current data across all 
meteorite groups.\ ArXiv e-prints arXiv:1408.6581. 

\item[--] Fogg, M.~J., Nelson, R.~P.\ 2005.\ Oligarchic and giant impact growth of terrestrial planets in the presence of gas giant planet migration.\ Astronomy and Astrophysics 441, 791-806. 

\item[--] Goldreich, P., 
Tremaine, S.\ 1980.\ Disk-satellite interactions.\ The Astrophysical 
Journal 241, 425-441. 

\item[--] Goldreich, P., 
Lithwick, Y., Sari, R.\ 2004.\ Final Stages of Planet Formation.\ The 
Astrophysical Journal 614, 497-507. 

\item[--] Greenberg, R., 
Hartmann, W.~K., Chapman, C.~R., Wacker, J.~F.\ 1978.\ Planetesimals to
planets - Numerical simulation of collisional evolution.\ {\it Icarus}
35, 1-26.

\item[--] Greenberg, R., 
Weidenschilling, S.~J., Chapman, C.~R., Davis, D.~R.\ 1984.\ From icy 
planetesimals to outer planets and comets.\ Icarus 59, 87-113.

\item[--] Greenzweig, 
Y., Lissauer, J.~J.\ 1990.\ Accretion rates of protoplanets.\ Icarus 87, 
40-77. 

\item[--] Greenzweig, 
Y., Lissauer, J.~J. 1992. Accretion rates of protoplanets. II - Gaussian 
distributions of planetesimal velocities.\ {\it Icarus} 100, 440-463.

\item[--] Guillot, T., Stevenson, D. J., Hubbard, W. B. and Saumon, D., 2004. The interior of Jupiter. in {\it Jupiter, the planet, satellites and magnetosphere}, Fran Bagenal, Timothy E. Dowling and William B. McKinnon Eds., The Cambridge University Press, 35--58.

\item [--] Guillot, T.\ 2005.\ The 
Interiors of Giant Planets: Models and Outstanding Questions.\ Annual 
Review of Earth and Planetary Sciences 33, 493-530. 

\item[--] Guillot, T., Ida, S., Ormel, C.~W.\ 2014.\ On the filtering and processing of dust by planetesimals. I. Derivation of collision probabilities for non-drifting planetesimals.\ Astronomy and Astrophysics 572, AA72. 

\item[--] G{\"u}ttler, C., 
Blum, J., Zsom, A., Ormel, C.~W., Dullemond, C.~P.\ 2009.\ The first phase 
of protoplanetary dust growth: The bouncing barrier.\ Geochimica et 
Cosmochimica Acta Supplement 73, 482. 

\item[--] Haisch, K.~E., Jr., 
Lada, E.~A., Lada, C.~J.\ 2001.\ Disk Frequencies and Lifetimes in Young 
Clusters.\ The Astrophysical Journal 553, L153-L156. 

\item[--] Halliday, A.~N., 
Kleine, T.\ 2006.\ Meteorites and the Timing, Mechanisms, and Conditions of 
Terrestrial Planet Accretion and Early Differentiation.\ Meteorites and the 
Early Solar System II 775-801. 

\item[--] Halliday, A.N. (2008) A young Moon-forming giant impact at 70-110 million years accompanied by late-stage mixing, core formation and degassing of the Earth. Royal Society of London Philosophical Transactions Series A 366, 4163-4181. 

\item[--] Hansen, B.~M.~S.\ 2009.\ 
Formation of the Terrestrial Planets from a Narrow Annulus.\ The 
Astrophysical Journal 703, 1131-1140. 

\item[--] Hayashi, C.\ 1981. Structure of the Solar
Nebula, Growth and Decay of Magnetic Fields and Effects of Magnetic
and Turbulent Viscosities on the Nebula.\ {\it Progress of Theoretical
Physics Supplement} 70, 35-53.

\item[--] Hartmann, L., Calvet, 
N., Gullbring, E., D'Alessio, P.\ 1998.\ Accretion and the Evolution of T 
Tauri Disks.\ The Astrophysical Journal 495, 385-400. 

\item[--] Hori, Y., Ikoma, M.\ 
2011.\ Gas giant formation with small cores triggered by envelope pollution 
by icy planetesimals.\ Monthly Notices of the Royal Astronomical Society 
416, 1419-1429. 

\item[--] Ikoma, M., Nakazawa, K., 
Emori, H.\ 2000.\ Formation of Giant Planets: Dependences on Core Accretion 
Rate and Grain Opacity.\ The Astrophysical Journal 537, 1013-1025. 

\item[--] Izidoro, A., 
Morbidelli, A., Raymond, S.~N.\ 2014.\ Terrestrial Planet Formation in the 
Presence of Migrating Super-Earths.\ The Astrophysical Journal 794, 11. 

\item[--] Izidoro, A., Raymond, 
S.~N., Morbidelli, A., Hersant, F., Pierens, A.\ 2015.\ Gas giant planets 
as dynamical barriers to inward-migrating super-Earths. In press in Astrophys. J. Lett.; \ ArXiv e-prints 
arXiv:1501.06308. 

\item[--] Jacobsen, S.~B.\ 2005.\ The 
Hf-W Isotopic System and the Origin of the Earth and Moon.\ Annual Review 
of Earth and Planetary Sciences 33, 531-570.

\item[--] Jacobson, S.~A., 
Morbidelli, A., Raymond, S.~N., O'Brien, D.~P., Walsh, K.~J., Rubie, D.~C.\ 
2014.\ Highly siderophile elements in Earth's mantle as a clock for the 
Moon-forming impact.\ Nature 508, 84-87. 

\item[--] Jacobson, 
S.~A., Morbidelli, A.\ 2014.\ Lunar and terrestrial planet formation in the 
Grand Tack scenario.\ Royal Society of London Philosophical Transactions 
Series A 372, 0174. 

\item[--] Johansen, A., 
Lacerda, P.\ 2010.\ Prograde rotation of protoplanets by accretion of 
pebbles in a gaseous environment.\ Monthly Notices of the Royal 
Astronomical Society 404, 475-485. 

\item[--] Johnson, T.~V., 
Lunine, J.~I.\ 2005.\ Saturn's moon Phoebe as a captured body from the 
outer Solar System.\ Nature 435, 69-71. 

\item[--] Johansen, A., Mac Low, 
M.-M., Lacerda, P., Bizzarro, M.\ 2015.\ Growth of asteroids, planetary 
embryos and Kuiper belt objects by chondrule accretion. Science Advances: e1500109 

\item[--] Klahr, H.~H., 
Bodenheimer, P.\ 2003.\ Turbulence in Accretion Disks: Vorticity Generation 
and Angular Momentum Transport via the Global Baroclinic Instability.\ The 
Astrophysical Journal 582, 869-892. 

\item[--] Kleine, T., Mezger, K., 
Palme, H., Scherer, E., M{\"u}nker, C.\ 2005.\ Early core formation in 
asteroids and late accretion of chondrite parent bodies: Evidence from 
$^{182}$Hf- $^{182}$W in CAIs, metal-rich chondrites, and iron meteorites.\ 
Geochimica et Cosmochimica Acta 69, 5805-5818. 

\item[--] Kokubo, E., Ida, S.\ 
1996.\ On Runaway Growth of Planetesimals.\ Icarus 123, 180-191. 

\item[--] Kokubo, E., Ida, S.\ 1998. 
Oligarchic Growth of Protoplanets.\ {\it Icarus} 131, 171-178.

\item[--] Kokubo, E., Ida, S.\ 2000.
Formation of Protoplanets from Planetesimals in the Solar Nebula.\ 
{\it Icarus} 143, 15-27.

\item[--] Kretke, K.~A., Lin, 
D.~N.~C.\ 2007.\ Grain Retention and Formation of Planetesimals near the 
Snow Line in MRI-driven Turbulent Protoplanetary Disks.\ The Astrophysical 
Journal 664, L55-L58. 

\item[--] Kretke, K.~A., 
Levison, H.~F.\ 2014a.\ Challenges in Forming the Solar System's Giant 
Planet Cores via Pebble Accretion.\ The Astronomical Journal 148, 109. 

\item[--] Kretke, K.~A., 
Levison, H.~F.\ 2014b.\ Forming Giant Planet Cores by Pebble Accretion -- 
Why Slow and Steady wins the Race.\ AAS/Division of Dynamical Astronomy 
Meeting 45, \#102.01. 

\item[--] Lambrechts, M., Johansen, A.\ 2012.\ Rapid growth of gas-giant cores by pebble accretion.\ Astronomy and Astrophysics 544, AA32. 

\item[--] Lambrechts, M., Johansen, A.\ 2014.\ Forming the cores of giant planets from the radial pebble flux in protoplanetary discs.\ Astronomy and Astrophysics 572, AA107. 

\item[--] Lambrechts, M., Johansen, A., Morbidelli, A.\ 2014.\ Separating gas-giant and ice-giant planets by halting pebble accretion.\ Astronomy and Astrophysics 572, AA35. 

\item[--] Levison, H.~F., 
Morbidelli, A.\ 2007.\ Models of the collisional damping scenario for 
ice-giant planets and Kuiper belt formation.\ Icarus 189, 196-212. 

\item[--] Levison, H.~F., 
Thommes, E., Duncan, M.~J.\ 2010.\ Modeling the Formation of Giant Planet 
Cores. I. Evaluating Key Processes.\ The Astronomical Journal 139, 
1297-1314. 

\item[--] Levison, H.~F., Duncan, 
M.~J., Minton, D.~M.\ 2012.\ Modeling Terrestrial Planet Formation with 
Full Dynamics,Accretion, and Fragmentation.\ AAS/Division for Planetary 
Sciences Meeting Abstracts 44, \#503.02. 

\item[--] Lissauer, J.~J.\ 1987.\ 
Timescales for planetary accretion and the structure of the protoplanetary 
disk.\ Icarus 69, 249-265. 

\item[--] Lodders, K.\ 2003.\ Solar 
System Abundances and Condensation Temperatures of the Elements.\ The 
Astrophysical Journal 591, 1220-1247. 

\item[--] Lynden-Bell, 
D., Pringle, J.~E.\ 1974.\ The evolution of viscous discs and the origin of 
the nebular variables..\ Monthly Notices of the Royal Astronomical Society 
168, 603-637. 

\item[--] Masset, F., 
Snellgrove, M.\ 2001.\ Reversing type II migration: resonance trapping of a 
lighter giant protoplanet.\ Monthly Notices of the Royal Astronomical 
Society 320, L55-L59. 

\item[--] Masset, F.~S., 
Casoli, J.\ 2009.\ On the Horseshoe Drag of a Low-Mass Planet. II. 
Migration in Adiabatic Disks.\ The Astrophysical Journal 703, 857-876. 

\item[--] Masset, F.~S., 
Casoli, J.\ 2010.\ Saturated Torque Formula for Planetary Migration in 
Viscous Disks with Thermal Diffusion: Recipe for Protoplanet Population 
Synthesis.\ The Astrophysical Journal 723, 1393-1417. 

\item[--] Mayor, M., and 13 
colleagues 2011.\ The HARPS search for southern extra-solar planets XXXIV. 
Occurrence, mass distribution and orbital properties of super-Earths and 
Neptune-mass planets.\ ArXiv e-prints arXiv:1109.2497. 

\item[--] McDonnell, J.~A.~M., Evans, G.~C., Evans, S.~T., Alexander, W.~M., Burton, W.~M., Firth, J.~G., Bussoletti, E., Grard, R.~J.~L., Hanner, M.~S., Sekanina, Z.\ 1987.\ The dust distribution within the inner coma of comet P/Halley 1982i - Encounter by Giotto's impact detectors.\ Astronomy and Astrophysics 187, 719-741. 

\item[--] Mizuno, H.\ 1980.\ Formation of 
the Giant Planets.\ Progress of Theoretical Physics 64, 544-557. 

\item[--] Morbidelli, A., 
Crida, A.\ 2007.\ The dynamics of Jupiter and Saturn in the gaseous 
protoplanetary disk.\ Icarus 191, 158-171. 

\item[--] Morbidelli, A., 
Bottke, W.~F., Nesvorn{\'y}, D., Levison, H.~F.\ 2009.\ Asteroids were born 
big.\ Icarus 204, 558-573. 

\item[--] Morbidelli, A., Nesvorny, D.\ 2012.\ Dynamics of pebbles in the vicinity of a growing planetary embryo: hydro-dynamical simulations.\ Astronomy and Astrophysics 546, AA18. 

\item[--] Morbidelli, A., 
Lunine, J.~I., O'Brien, D.~P., Raymond, S.~N., Walsh, K.~J.\ 2012.\ 
Building Terrestrial Planets.\ Annual Review of Earth and Planetary 
Sciences 40, 251-275.

\item[--] Morfill, G.~E., 
Voelk, H.~J.\ 1984.\ Transport of dust and vapor and chemical fractionation 
in the early protosolar cloud.\ The Astrophysical Journal 287, 371-395. 

\item[--] Movshovitz, N., 
Podolak, M.\ 2008.\ The opacity of grains in protoplanetary atmospheres.\ 
Icarus 194, 368-378. 

\item[--]  Murray-Clay, R., 
Kratter, K., \& Youdin, A.\ 2011.\ AAS/Division for Extreme Solar Systems Abstracts 2, 804. 

\item[--] Nelson, R.~P., Gressel, 
O., Umurhan, O.~M.\ 2013.\ Linear and non-linear evolution of the vertical 
shear instability in accretion discs.\ Monthly Notices of the Royal 
Astronomical Society 435, 2610-2632. 

\item[--] Nettelmann, N., 
Holst, B., Kietzmann, A., French, M., Redmer, R., Blaschke, D.\ 2008.\ Ab 
Initio Equation of State Data for Hydrogen, Helium, and Water and the 
Internal Structure of Jupiter.\ The Astrophysical Journal 683, 1217-1228. 

\item[--] O'Brien, D.~P., 
Morbidelli, A., Levison, H.~F.\ 2006.\ Terrestrial planet formation with 
strong dynamical friction.\ Icarus 184, 39-58. 

\item[--] Ormel, C.~W., Klahr, H.~H.\ 2010.\ The effect of gas drag on the growth of protoplanets. Analytical expressions for the accretion of small bodies in laminar disks.\ Astronomy and Astrophysics 520, AA43. 

\item[--] Ormel, C.~W.\ 2014.\ An 
Atmospheric Structure Equation for Grain Growth.\ The Astrophysical Journal 
789, L18. 

\item[--] Paardekooper, S.-J., Mellema, G.\ 2006.\ Halting type I planet migration in non-isothermal disks.\ Astronomy and Astrophysics 459, L17-L20. 

\item[--] Paardekooper, 
S.-J., Baruteau, C., Crida, A., Kley, W.\ 2010.\ A torque formula for 
non-isothermal type I planetary migration - I. Unsaturated horseshoe drag.\ 
Monthly Notices of the Royal Astronomical Society 401, 1950-1964. 

\item[--] Paardekooper, 
S.-J., Baruteau, C., Kley, W.\ 2011.\ A torque formula for non-isothermal 
Type I planetary migration - II. Effects of diffusion.\ Monthly Notices of 
the Royal Astronomical Society 410, 293-303. 

\item[--] Pollack, J.~B., 
Hubickyj, O., Bodenheimer, P., Lissauer, J.~J., Podolak, M., Greenzweig, 
Y.\ 1996.\ Formation of the Giant Planets by Concurrent Accretion of Solids 
and Gas.\ Icarus 124, 62-85. 

\item[--] Raymond, S.~N., Quinn, 
T., Lunine, J.~I.\ 2004.\ Making other earths: dynamical simulations of 
terrestrial planet formation and water delivery.\ Icarus 168, 1-17. 

\item[--] Raymond, S.~N., Quinn, 
T., Lunine, J.~I.\ 2006a.\ High-resolution simulations of the final assembly 
of Earth-like planets I. Terrestrial accretion and dynamics.\ Icarus 183, 
265-282. 

\item[--] Raymond, S.~N., 
Mandell, A.~M., Sigurdsson, S.\ 2006b.\ Exotic Earths: Forming Habitable 
Worlds with Giant Planet Migration.\ Science 313, 1413-1416. 

\item[--] Ros, K., Johansen, A.\ 2013.\ Ice condensation as a planet formation mechanism.\ Astronomy and Astrophysics 552, AA137. 

\item[--] Shakura, N.~I., Sunyaev, R.~A.\ 1973.\ Black holes in binary systems. Observational appearance..\ Astronomy and Astrophysics 24, 337-355. 

\item[--] Stern, S.~A., McKinnon, 
W.~B., Lunine, J.~I.\ 1997.\ On the Origin of Pluto, Charon, and the 
Pluto-Charon Binary.\ Pluto and Charon 605. 

\item[--] Stevenson, D.~J.\ 1982.\ Formation of the giant planets.\ Planetary and Space Science 30, 755-764. 

\item[--] Stewart, G.~R., Ida, 
S.\ 2000.\ Velocity Evolution of Planetesimals: Unified Analytical Formulas 
and Comparisons with N-Body Simulations.\ Icarus 143, 28-44. 

\item[--] Stoll, M.~H.~R., Kley, W.\ 2014.\ Vertical shear instability in accretion disc models with radiation transport.\ Astronomy and Astrophysics 572, A77. 

\item[--] Stone, J.~M., Gammie, 
C.~F., Balbus, S.~A., Hawley, J.~F.\ 2000.\ Transport Processes in 
Protostellar Disks.\ Protostars and Planets IV 589. 

\item[--] Tanaka, H., Takeuchi, 
T., Ward, W.~R.\ 2002.\ Three-Dimensional Interaction between a Planet and 
an Isothermal Gaseous Disk. I. Corotation and Lindblad Torques and Planet 
Migration.\ The Astrophysical Journal 565, 1257-1274. 

\item[--] Taylor, D.J., McKeegan, K.D., Harrison, T.M., Young, E.D. (2009) Early differentiation of the lunar magma ocean. New Lu-Hf isotope results from Apollo 17. Geochimica et Cosmochimica Acta Supplement 73, 1317. 

\item[--] Thommes, E.~W., Duncan, 
M.~J., Levison, H.~F.\ 2003.\ Oligarchic growth of giant planets.\ Icarus 
161, 431-455. 

\item[--] Touboul M, Kleine T, Bourdon B, Palme H, Wieler R (2007) Late formation and prolonged differentiation of the Moon inferred from W isotopes in lunar metals. Nature 450:1206-1209 

\item[--] Turner, N.~J., Sano, T., 
Dziourkevitch, N.\ 2007.\ Turbulent Mixing and the Dead Zone in 
Protostellar Disks.\ The Astrophysical Journal 659, 729-737. 

\item[--] Walsh, K.~J., Morbidelli, 
A., Raymond, S.~N., O'Brien, D.~P., Mandell, A.~M.\ 2011.\ A low mass for 
Mars from Jupiter's early gas-driven migration.\ Nature 475, 206-209. 

\item[--] Weidenschilling, S.~J.\ 1977. The
 distribution of mass in the planetary system and solar nebula.\
 {\it Astrophysics and Space Science} 51, 153-158.

\item[--] Weidenschilling, S.~J., Spaute, D., Davis, D.~R., Marzari, F., Ohtsuki, K.\ 1997. Accretional
 Evolution of a Planetesimal Swarm.\ {\it Icarus} 128, 429-455.

\item[--] Wetherill,
G.~W., Stewart, G.~R.\ 1989. Accumulation of a swarm of small
planetesimals.\ {\it Icarus} 77, 330-357.

\item[--] Wetherill, G.~W.\ 1992.\ An 
alternative model for the formation of the asteroids.\ Icarus 100, 307-325. 

\item[--] Wetherill,
G.~W., Stewart, G.~R. 1993. Formation of planetary embryos - Effects of
fragmentation, low relative velocity, and independent variation of
eccentricity and inclination.\ {\it Icarus} 106, 190-205.

\item[--] Wilson, H.~F., 
Militzer, B.\ 2012.\ Solubility of Water Ice in Metallic Hydrogen: 
Consequences for Core Erosion in Gas Giant Planets.\ The Astrophysical 
Journal 745, 54. 

\item[--] Yin Q.Z., Jacobsen SB, Yamashita K, Blichert-Toft J, Telouk P, Albarede F (2002) A short timescale for terrestrial planet formation from Hf-W chronometry of meteorites. Nature 418(6901):949-952 

\item[--] Youdin, A.~N., 
Lithwick, Y.\ 2007.\ Particle stirring in turbulent gas disks: Including 
orbital oscillations.\ Icarus 192, 588-604. 

\item[--] Wada, K., Tanaka, H., 
Suyama, T., Kimura, H., Yamamoto, T.\ 2009.\ Collisional Growth Conditions 
for Dust Aggregates.\ The Astrophysical Journal 702, 1490-1501. 

\end{itemize}

%\end{thebibliography}
\end{document}